\documentclass[twocolumn,10pt,superscriptaddress,amsmath,amssymb,aps,pra,floatfix,]{revtex4-2}
\usepackage{graphicx}
\usepackage{bm}
\usepackage{times}
\usepackage{amsmath,amssymb}
\usepackage{mathtools}
\usepackage{enumerate}
\usepackage{multirow}
\usepackage{xcolor}
\usepackage{braket}
 
\usepackage[colorlinks=true,linkcolor=blue,citecolor=blue, linktocpage=true,breaklinks=true]{hyperref}

\newcommand{\eq}{\begin{equation}}
\newcommand{\en}{\end{equation}}
\newcommand{\eqa}{\begin{eqnarray}}
\newcommand{\ena}{\end{eqnarray}}
\newcommand{\tr}{\mathrm{Tr}}

\begin{document}

\title{Ergotropy Activation and Coherence in Dissipatively Charged Quantum Batteries}

\author{Hui-Yu Yang}
\affiliation{School of Physics, Northwest University, Xi'an 710127, China}

\author{Kang-Yuan Chen}
\affiliation{School of Physics, Northwest University, Xi'an 710127, China}

\author{Kun Zhang}
\affiliation{School of Physics, Northwest University, Xi'an 710127, China}
\affiliation{Shaanxi Key Laboratory for Theoretical Physics Frontiers, Xi'an 710127, China}
\affiliation{Peng Huanwu Center for Fundamental Theory, Xi'an 710127, China}
\affiliation{Fundamental Discipline Research Center for  Quantum Science and technology of Shaanxi Province, Xi'an 710127, China}

\author{Yang-Yang Chen}
\affiliation{Shaanxi Key Laboratory for Theoretical Physics Frontiers, Xi'an 710127, China}
\affiliation{Peng Huanwu Center for Fundamental Theory, Xi'an 710127, China}
\affiliation{Fundamental Discipline Research Center for  Quantum Science and technology of Shaanxi Province, Xi'an 710127, China}
\affiliation{Institute of Modern Physics, Northwest University, Xi'an 710127, China}

\author{Xiao-Hui Wang}
\email{xhwang@nwu.edu.cn}
\affiliation{School of Physics, Northwest University, Xi'an 710127, China}
\affiliation{Shaanxi Key Laboratory for Theoretical Physics Frontiers, Xi'an 710127, China}
\affiliation{Peng Huanwu Center for Fundamental Theory, Xi'an 710127, China}
\affiliation{Fundamental Discipline Research Center for  Quantum Science and technology of Shaanxi Province, Xi'an 710127, China}


\begin{abstract}
Quantum batteries can exploit quantum resources to enhance energy storage and work extraction, and recent experimental progress has brought their implementation closer to reality. 
In realistic settings, however, environmental dissipation is unavoidable, making dissipative charging a physically relevant operating regime. 
Here we investigate ergotropy in dissipatively charged quantum batteries, focusing on interaction-induced work extraction. 
For two coupled two-level systems (TLSs), we derive exact analytical expressions for the total, coherent, and incoherent ergotropy in both resonant and detuned configurations. 
We show that, in the resonant case, ergotropy is activated only above a critical coupling associated with an interaction-induced energy-level crossing, whereas detuning enables ergotropy already in the weak-coupling regime. 
Moreover, quantum coherence, although necessary for coherent work extraction, is not by itself sufficient to guarantee nonzero coherent ergotropy. 
In the strong-coupling regime, the resonant setup reaches the maximum ergotropy at a finite coupling strength, whereas the detuned setup approaches the same limit only asymptotically. 
Extending the analysis to a periodic XX spin chain of $N$ TLSs, we analytically derive the finite-temperature coupling threshold for ergotropy activation and show that it scales as $1/N$ in the large-$N$ limit.
These results identify frequency detuning, interaction strength, and system size as key control parameters for ergotropy generation in open quantum batteries.
\end{abstract}

\maketitle

\section{\label{sec:intro} Introduction}

Quantum batteries have attracted broad interest as a platform for exploring quantum-enhanced energy storage and work extraction \cite{campaioli2024colloquium,PhysRevE.87.042123,PhysRevLett.111.240401,PhysRevA.110.052404,PhysRevLett.131.030402,PhysRevResearch.2.023113}. 
Compared with their classical counterparts, they can exploit genuinely quantum resources, especially entanglement and coherence, to enhance storage capacity, charging power, and operational performance \cite{PhysRevA.110.062209,PhysRevLett.127.100601}. 
At the same time, rapid progress in superconducting-qubit platforms has brought these ideas closer to experimental implementation, while also making clear that environmental noise, decoherence, and energy dissipation are unavoidable ingredients in realistic charging protocols \cite{barra2019dissipative,mayo2022collective,zakavati2021bounds,lu2025topological}.

A central figure of merit for quantum batteries is the amount of useful work that can be extracted from the charged state, which is quantified by ergotropy \cite{allahverdyan2004maximal}. 
By definition, ergotropy is the maximum work that can be extracted from a quantum state through unitary operations. 
When evaluated in the energy eigenbasis relevant to the work extraction stage, it can be further decomposed into coherent and incoherent contributions, which distinguish the roles of quantum superpositions and population redistribution, respectively. 

Recent studies have shown that coherence is an important resource for work extraction \cite{korzekwa2016extraction, hadipour2024work,plesnik2024impossibility,lobejko2022work,tirone2025quantum,chen2025work,williamson2024extracting}. 
In particular, fundamental bounds have been established between coherent ergotropy and quantum coherence \cite{PhysRevLett.125.180603}, and subsequent studies further clarified that incoherent ergotropy is closely related to diagonal entropy, whereas coherent ergotropy is directly linked to quantum coherence \cite{Shi2022Entanglement}. 
\begingroup

Entanglement alone cannot account for enhanced quantum battery performance, whereas coherence can improve the charging process~\cite{Kamin2020}. More generally, coherent ergotropy depends jointly on coherence and the participation ratio, whereas population inversion enables an additional incoherent contribution~\cite{WangDou2026}.
\endgroup
\hypersetup{linkcolor=blue,citecolor=blue}
Related investigations in open quantum systems have also shown that coherent ergotropy can display enhanced robustness against environmental decoherence \cite{song2025self}, while spectrum engineering, for instance through topological frustration, can help protect coherence-related battery performance \cite{catalano2024frustrating}. 
Taken together, these results establish ergotropy and its coherent-incoherent decomposition as a natural framework for understanding how quantum resources contribute to battery performance.

Most existing quantum battery protocols can be broadly classified into two charging paradigms. 
One is based on auxiliary quantum chargers, as in the Tavis--Cummings model \cite{PhysRevA.109.012204,fink2009dressed,tessier2003entanglement}, the central-spin model \cite{peng2021lower,liu2021entanglement,yang2025optimal}, and the Dicke model \cite{roses2020dicke,ferraro2018high,yang2024three}. 
The other relies on externally driven schemes, such as the Sachdev--Ye--Kitaev model \cite{rossini2020quantum,PhysRevA.110.062209} and spin-chain protocols \cite{le2018spin,dou2022cavity}.
In realistic implementations, however, the battery is inevitably coupled to dissipative environments, which can strongly affect both charging and storage performance. 
Although approaches such as reservoir engineering \cite{song2024remote}, Floquet control \cite{bai2020floquet}, feedback protocols \cite{yao2022optimal}, and dark-state protection \cite{quach2020using} can mitigate decoherence, dissipation is still often regarded primarily as a limitation.

Recent work has instead introduced a dissipative charging paradigm in which internal interactions are suddenly switched on to drive the battery out of local equilibrium, and are then quenched off so that the charging cycle is completed with respect to the bare battery Hamiltonian. 
This mechanism can produce a collective enhancement of charging power that scales as $N^{2}$ \cite{pokhrel2025large}. 
However, enhanced charging power does not by itself determine how much useful work can ultimately be extracted from the final charged state, and the corresponding ergotropy limit has not yet been systematically quantified. 
Since dissipation alone cannot charge the battery, controlled internal interactions are essential for activating and amplifying ergotropy. 
Meanwhile, coherence may persist even in thermal states, giving rise to what is commonly called thermal coherence \cite{ccakmak2017thermal,parthasarathy2021thermal,li2022thermal,ferreira2023thermal,wang2019steady}, suggesting a possible resource for work extraction. 
A central question therefore remains: how do thermal coherence, interaction-induced energy-level crossings, and frequency detuning jointly govern ergotropy generation in dissipatively charged multi-qubit batteries?

In this work, we investigate ergotropy generation in dissipatively charged quantum batteries driven by internal interactions. 
\begingroup

Here we reveal the mechanisms and analytical thresholds of ergotropy activation in the dissipative charging protocol and show how they are controlled by interaction strength, frequency detuning, and system size.
\endgroup
Starting from a two-qubit model in both resonant (qubits with equal frequencies) and detuned (qubits with different frequencies) configurations, we show that the extractable work is governed by the interplay between frequency detuning, interaction strength, and interaction-induced energy-level crossings. 
Our results further show that, although coherence is necessary for coherent ergotropy, it is not by itself sufficient to guarantee nonzero coherent extractable work. 
We also uncover a clear operational asymmetry between the two charging schemes: detuning lowers the activation threshold for ergotropy generation, whereas the resonant configuration is more effective in driving the extractable work toward its maximal value. 
Finally, by extending the analysis to an $N$-spin XX chain, we show that increasing system size promotes interaction-induced level crossings and thereby facilitates ergotropy generation. 
Taken together, these results identify frequency detuning, interaction strength, and system size as key control parameters for optimizing work extraction in dissipative quantum batteries.

The remainder of this paper is organized as follows. Sec.~\ref{preliminary} introduces the model and the ergotropy. Sec.~\ref{two TLS} presents the analytical results for two coupled TLSs. Secs.~\ref{balanced charging scheme} and \ref{detuning Charging scheme} examine the resonant and detuned charging schemes, respectively. Sec.~\ref{XX spin} extends the analysis to the multi-spin XX chain. The last section is the conclusion.

\begin{figure}[t]
\includegraphics[width=1\columnwidth]{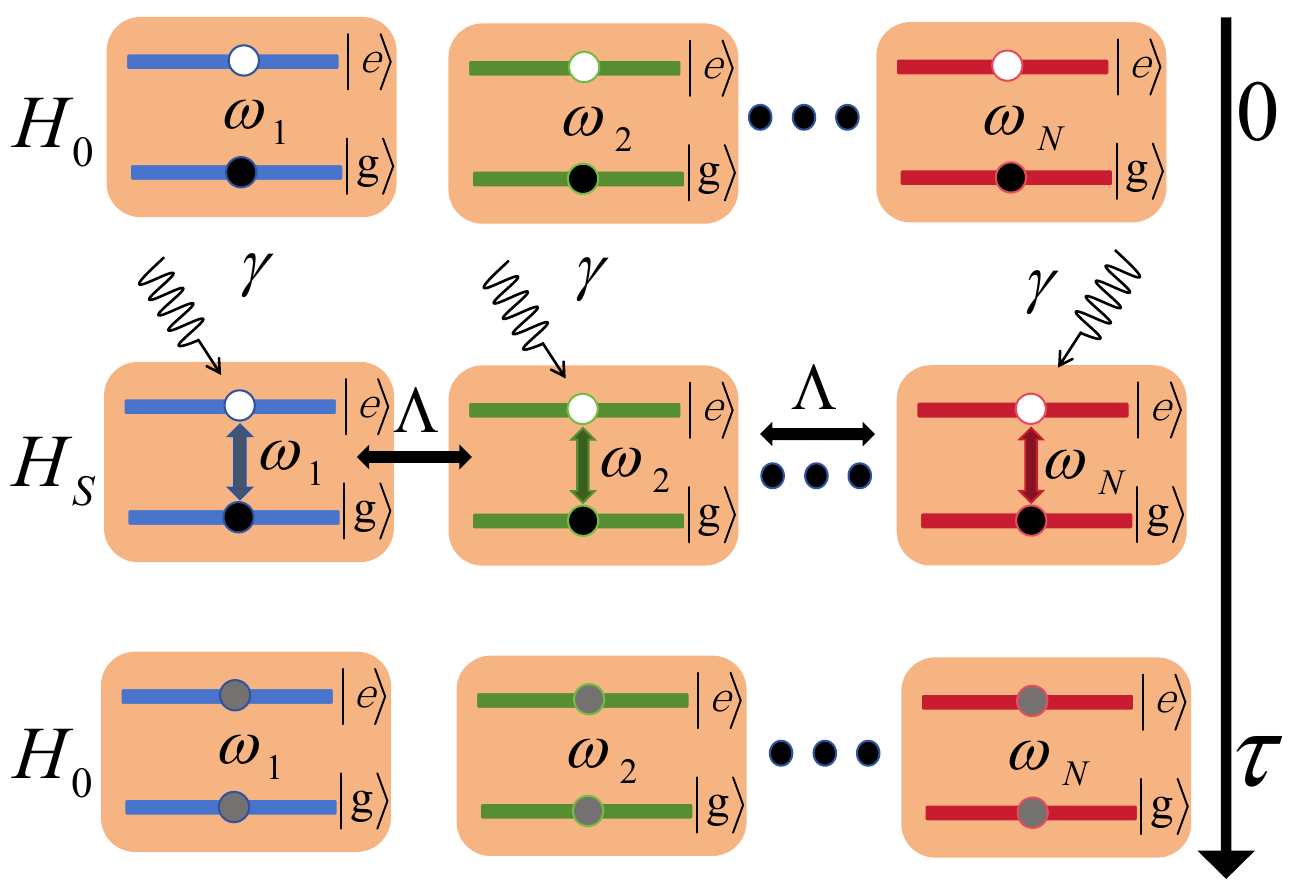}
\caption{Dissipative charging protocol. Here $H_{0}$ denotes the Hamiltonian of the $N$ TLSs battery, and $H_{S}$ is the Hamiltonian when the interaction term $H_{I}$ is switched on. For $t<0$, the battery is prepared in its ground state. During the charging stage, $0<t<\tau$, the interaction $H_I$ is activated and the state relaxes toward the thermal state under dissipative dynamics. At $t=\tau$, the charging Hamiltonian $H_{I}$ is switched off, and energy is stored in the battery.}\label{fig1}
\end{figure}

\section{Preliminaries}\label{preliminary}

In this section, we set up the model and introduce the basic notation for ergotropy. We begin in Sec.~\ref{model} by defining the model and the dissipative charging protocol. In Sec.~\ref{sec:QBperformance}, we define ergotropy and its coherent and incoherent components, which together provide a comprehensive characterization of battery performance.

\subsection{Model and dissipative charging protocol}\label{model}

We consider a quantum battery composed of $N$ two-level systems (TLSs), described by the Hamiltonian
\begin{subequations}
\begin{align}
H_S &= H_0 + g(t) H_I, \\
H_0 &= \sum_{j=1}^{N} \omega_j \sigma_j^z, \\
H_I &= \frac{\Lambda}{2} \sum_{j=1}^{N} \left(\sigma_j^x \sigma_{j+1}^x + \sigma_j^y \sigma_{j+1}^y\right),
\end{align}   
\end{subequations}
with periodic boundary conditions $\sigma_{N+1}=\sigma_1$. Here $H_0$ denotes the bare battery Hamiltonian, with $\omega_j$ the onsite energies of the battery spins, and $H_I$ describes the flip-flop interaction with coupling strength $\Lambda$. The switch function $g(t)$ controls the charging protocol: $g(t)=1$ for $t\in[0,\tau]$ and $g(t)=0$ otherwise, where $\tau$ is the charging time.

\begingroup

During charging ($0<t<\tau$), each TLS is weakly coupled to an independent bosonic reservoir at inverse temperature $\beta$. At $t=0$, the battery is initialized in the ground state of $H_0$, $\rho(0)=|1\rangle\langle1|^{\otimes N}$, with $\sigma_j^z|1\rangle=-|1\rangle$, and $H_I$ is switched on. The microscopic model is described by $H_{\mathrm{tot}}=H_S+H_B+H_{SB}$, where $H_B$ describes the bosonic reservoirs
\begin{equation}
H_B=\sum_{j,k}\Omega_{jk}b_{jk}^{\dagger}b_{jk},
\end{equation}
and $H_{SB}$ describes the battery--reservoir interaction
\begin{equation}
H_{SB}=\sum_{j,k}\lambda_{jk}\sigma_j^x\otimes\left(b_{jk}+b_{jk}^{\dagger}\right).
\end{equation}
Here $b_{jk}$ annihilates mode $k$ of reservoir $j$ with frequency $\Omega_{jk}$, and $\lambda_{jk}$ is the corresponding real coupling amplitude. 

Under the Born--Markov and secular approximations in the global eigenbasis of the interacting Hamiltonian $H_S$, the bath-induced rates obey the detailed-balance relation $\gamma_j(-\nu)=e^{-\beta\nu}\gamma_j(\nu)$, which ensures that the Gibbs state is stationary~\cite{Breuer2002,Scala2008}. Assuming that the dissipative dynamics has a unique stationary state, the system relaxes to the Gibbs equilibrium state, given by
\begin{equation}
\rho = \frac{e^{-\beta H_S}}{Z}, \qquad Z=\tr\!\left(e^{-\beta H_S}\right).
\end{equation} 
Denote the relaxation time by $\tau_{\mathrm{rel}}$. We then switch off the battery--reservoir interaction at $t=\tau$, with $\tau\gg\tau_{\mathrm{rel}}$. Exact degeneracies are treated by grouping transitions with the same Bohr frequency. When distinct Bohr frequencies become comparable to dissipative linewidths, a partial-secular or unified weak-coupling treatment may instead be required~\cite{Cattaneo2020,Trushechkin2021}. The zero-temperature results at a level crossing are understood as the limit $\beta\to\infty$ taken after finite-temperature equilibration.
\endgroup
\hypersetup{linkcolor=blue,citecolor=blue}

After charging, the interaction term $H_I$ is quenched off, and the stored energy and extractable work are evaluated with respect to the bare Hamiltonian $H_0$. In this sense, the charging stage prepares a thermal state of $H_S$, whereas the subsequent work extraction stage is defined relative to $H_0$. The present charging protocol can be regarded as an effective and experimentally relevant open spin-$1/2$ model. The bare term $H_0$ captures locally tunable level splittings, while $H_I$ represents exchange-type couplings that can be engineered in several quantum-simulation platforms, including superconducting qubits and Rydberg-atom arrays \cite{Nguyen2024,Zhou2021,Lippe2021}. The dissipative charging stage is compatible with setups employing controlled relaxation or engineered reservoirs. From this perspective, the model provides a minimal framework for studying the dissipatively charged quantum battery. 

Moreover, the present dissipative charging protocol is closely related in spirit to direct charging schemes based on a sudden quench of the interaction Hamiltonian. The key difference is that here the coupling to the environment is explicitly incorporated during the charging stage, so that, in the long-time limit, the battery is prepared in the Gibbs state of the interacting Hamiltonian $H_S$ rather than in a purely unitary nonequilibrium state. In this sense, the protocol provides a natural open-system extension of quench-based charging and incorporates dissipation into the charging process.

\subsection{Battery performance: Ergotropy}\label{sec:QBperformance}

The central metric for characterizing quantum battery performance is the stored energy
\begin{equation}
E(\rho) = \tr(H_0 \rho),
\end{equation}
evaluated with respect to the bare battery Hamiltonian $H_0$. In general, only a part of this stored energy can be extracted as work by unitary operations. The corresponding maximum extractable work is quantified by the ergotropy $\mathcal{E}(\rho)$~\cite{allahverdyan2004maximal},
\begin{equation}
\mathcal{E}(\rho) = E(\rho) - E(\rho_p),
\end{equation}
\begingroup

operationally, discharging means applying the optimal unitary operation that maps the charged state $\rho$ onto $\rho_p$ and extracts the maximum work $\mathcal{E}(\rho)$,
\endgroup
where $\rho_p$ is the passive state associated with $\rho$. It is constructed by assigning the eigenvalues of $\rho$ in nonincreasing order to the eigenstates of $H_0$ in nondecreasing energy order. Explicitly, if $\rho = \sum_k p_k \ket{p_k}\bra{p_k}$ with $p_k \ge p_{k+1}$ and $H_0 = \sum_k \epsilon_k \ket{\epsilon_k}\bra{\epsilon_k}$ with $\epsilon_k \le \epsilon_{k+1}$, the passive state is given by $\rho_p = \sum_k p_k \ket{\epsilon_k}\bra{\epsilon_k}$.

Since the work extraction stage is defined relative to $H_0$, the coherent and incoherent contributions to ergotropy are also defined with respect to the eigenbasis of $H_0$~\cite{PhysRevLett.125.180603,Shi2022Entanglement,song2025self}. To isolate the population contribution, we dephase $\rho$ in the eigenbasis of $H_0$, denoted as
\begin{equation}
\pi = \Delta[\rho] = \sum_k \bra{\epsilon_k}\rho\ket{\epsilon_k}\ket{\epsilon_k}\bra{\epsilon_k}.
\end{equation}
Here $\pi$ has the same diagonal populations as $\rho$ in the eigenbasis of $H_0$, but no coherence (off-diagonal term) in the eigenbasis $|\epsilon_k\rangle$. The incoherent ergotropy is then defined as
\begin{equation}
\label{eq:E_i}
\mathcal{E}_i(\rho) = E(\pi) - E(\pi_p),
\end{equation}
where $\pi_p$ is the passive state associated with $\pi$. The coherent contribution to ergotropy is defined as the difference between the total and incoherent parts,
\begin{equation}
\mathcal{E}_c(\rho) = \mathcal{E}(\rho) - \mathcal{E}_i(\rho).
\end{equation}
Thus, $\mathcal{E}_c(\rho)$ quantifies the part of the extractable work that originates from coherence in the eigenbasis of $H_0$.

To quantify this coherence, we use the relative entropy of coherence~\cite{baumgratz2014quantifying},
\begin{equation}
\label{C}
C(\rho) = S(\pi) - S(\rho),
\end{equation}
where $S(\rho) = -\tr(\rho \log \rho)$ is the von Neumann entropy. Here $C(\rho)$ measures the coherence of $\rho$ in the eigenbasis of $H_0$, which is the relevant basis for the coherent-incoherent decomposition of ergotropy in our study.

\section{Analytical Analysis: Two TLS\MakeLowercase{s}}\label{two TLS}

In this section, we consider the quantum battery composed of two TLSs and coupled to a dissipative environment, with charging induced by the two-qubit interactions.
We begin in Sec.~\ref{Thermal State} by determining the spectral properties of the interacting Hamiltonian and, accordingly, obtaining the exact thermal state.
This formulation then enables us to derive exact analytical expressions for the total ergotropy in Sec.~\ref{total ergotropy}, as well as its incoherent and coherent contributions in Secs.~\ref{Incoherent ergotropy} and \ref{coherent ergotropy}, respectively.

\subsection{Spectral decomposition and thermal state}\label{Thermal State}
To determine the thermal state obtained at the end of the charging stage, we first derive the exact eigenspectrum of the charging battery Hamiltonian $H_S$. This is achieved by transforming from the local computational basis to the dressed-state basis through a unitary rotation. Specifically, introducing the mixing angle
\begin{equation}
\label{eq:theta}
\theta=\tan^{-1}\!\left(\frac{\kappa-\delta}{2\Lambda}\right), \qquad 0\le\theta\le\pi,
\end{equation}
with $\delta=\omega_1-\omega_2$ and $\kappa=\sqrt{4\Lambda^2+\delta^2}$, the Hamiltonian is diagonalized as
\begin{align}\label{Hs}
\widetilde{H}_{S}=e^{iG\theta}H_S e^{-iG\theta}
=\mathrm{diag}(\omega,\kappa,-\kappa,-\omega),
\end{align}
where $\omega=\omega_1+\omega_2$ and
\begin{equation}
G=\frac{1}{2}\left(\sigma^y_1\sigma^x_2-\sigma^x_1\sigma^y_2\right).
\end{equation}
The corresponding eigenvalues and eigenvectors are
\begin{subequations}
\label{eq:eigenbasis_eigenvalues}
\begin{alignat}{3}
E_1 &= -\omega, &\quad |E_1\rangle &= |11\rangle, \\
E_2 &= -\kappa, &\quad |E_2\rangle &= \cos\theta\,|01\rangle + \sin\theta\,|10\rangle, \\
E_3 &= \kappa, &\quad |E_3\rangle &= \sin\theta\,|01\rangle - \cos\theta\,|10\rangle, \\
E_4 &= \omega, &\quad |E_4\rangle &= |00\rangle,
\end{alignat}
\end{subequations}
where the angle is given by Eq.~\eqref{eq:theta}. Accordingly, the thermal state in the dressed basis is
\begin{equation}
\widetilde{\rho}_{S}
=\frac{1}{Z}\mathrm{diag}\left(e^{-\beta\omega},\,e^{-\beta\kappa},\,e^{\beta\kappa},\,e^{\beta\omega}\right),
\end{equation}
with $Z=2\left(\cosh(\beta\kappa)+\cosh(\beta\omega)\right)$. Transforming back to the local computational basis, we obtain
\begin{subequations}
\label{rho}
\begin{align}
\rho^{11}&=\frac{e^{-\beta\omega}}{Z}, \qquad
\rho^{44}=\frac{e^{\beta\omega}}{Z}, \\
\rho^{22}&=\frac{1}{Z}\left(\cosh(\beta\kappa)-\frac{\delta}{\kappa}\sinh(\beta\kappa)\right), \\
\rho^{23}&=\rho^{32}=-\frac{2\Lambda}{\kappa Z}\sinh(\beta\kappa), \\
\rho^{33}&=\frac{1}{Z}\left(\cosh(\beta\kappa)+\frac{\delta}{\kappa}\sinh(\beta\kappa)\right).
\end{align}
\end{subequations}
\begingroup

The Gibbs state of $H_S$ generally contains nonzero off-diagonal terms in the local product basis, which is an eigenbasis of $H_0$, even though it is diagonal in the eigenbasis of $H_S$. Such interaction-induced equilibrium coherences are commonly referred to as thermal coherence~\cite{ccakmak2017thermal,parthasarathy2021thermal,li2022thermal,ferreira2023thermal,wang2019steady}. 

The energetic role of the local-basis coherence depends on the detuning. For $\delta\neq0$, one has $[H_0,H_I]\neq0$, so the eigenbases of $H_S$ and $H_0$ are misaligned and the resulting coherence can connect distinct bare-energy levels and contribute to ergotropy. At resonance ($\delta=0$), however, $[H_0,H_I]=0$. The interaction mixes states only within degenerate eigenspaces of $H_0$. Off-diagonal terms can therefore appear in the local product basis without connecting distinct bare-energy levels. This distinction underlies the different ergotropy-activation mechanisms in the resonant and detuned cases discussed separately in Secs. \ref{balanced charging scheme} and \ref{detuning Charging scheme} respectively. 
\endgroup
\hypersetup{linkcolor=blue,citecolor=blue}

\begin{figure}[t]
\includegraphics[width=3.3in]{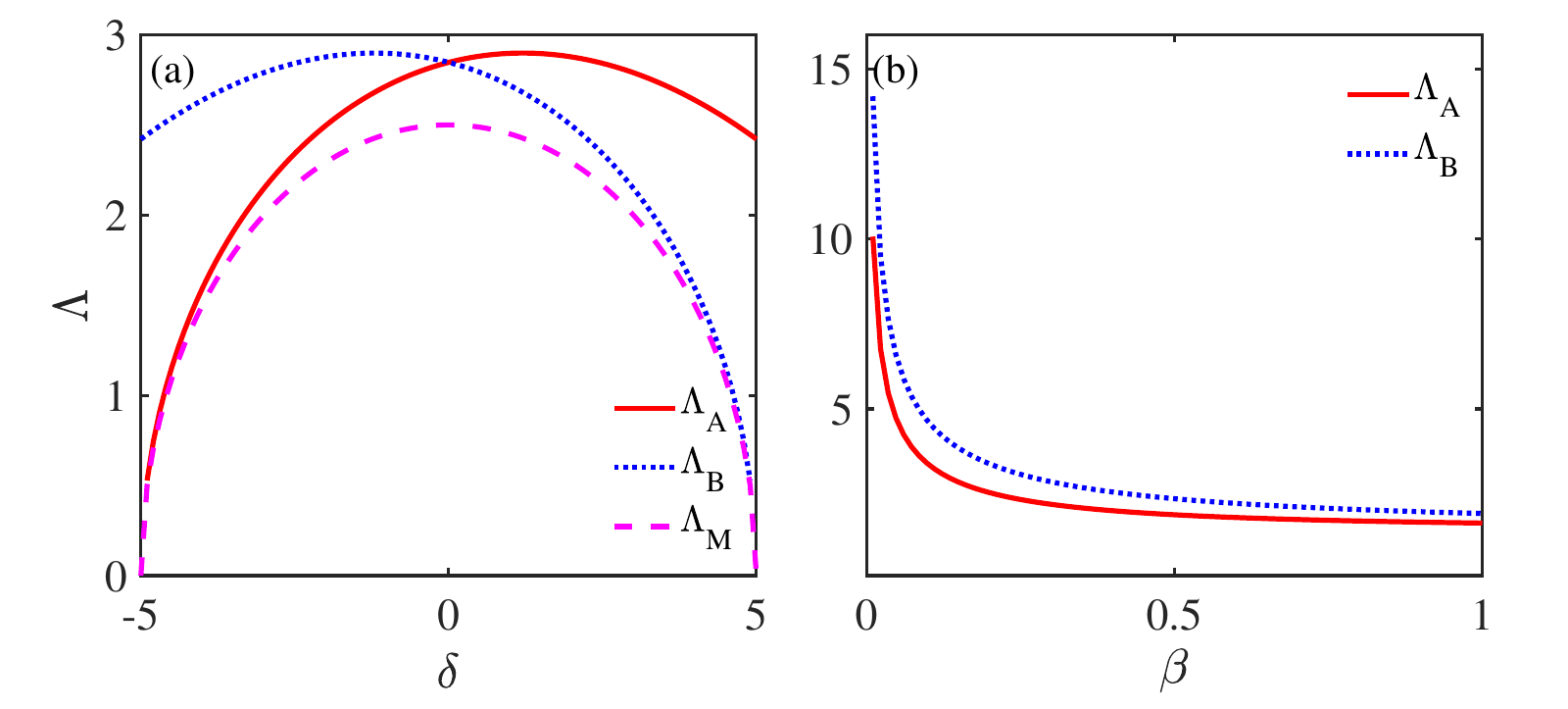}
\caption{Ordering of the critical coupling strengths $\Lambda_M$, $\Lambda_A$, and $\Lambda_B$. (a) Dependence of $\Lambda_M$, $\Lambda_A$, and $\Lambda_B$ on the detuning $\delta=\omega_1-\omega_2$ for $\omega=\omega_1+\omega_2=5$ and $\beta=1$. (b) Dependence of $\Lambda_A$ and $\Lambda_B$ on the inverse temperature $\beta$ for $\omega_1=1$ and $\omega_2=2$.}\label{98}
\end{figure}

\subsection{Total ergotropy}\label{total ergotropy}
The energy stored in the battery after the charging stage is
\begin{equation}
E(\rho)=\tr(\rho H_0 )=-\frac{2}{\kappa Z} \left(\delta ^2 \sinh (\beta \kappa )+\kappa \omega \sinh (\beta \omega )\right).
\end{equation}
The corresponding passive state $\rho_p$ depends on the ordering between the populations of $\rho$ and the eigenenergies of $H_0$.
A key change occurs at the coupling strength
\begin{align}\label{Delta}
\Lambda _M=\sqrt{\omega_1 \omega_2},
\end{align}
which marks the level crossing in the spectrum of $H_S$ between $-\omega$ and $-\kappa$. For $\Lambda < \Lambda_M$, as shown in Eq. \eqref{eq:eigenbasis_eigenvalues}, the system lies in the weak-coupling regime, where the ground state remains the separable state $|11\rangle$. For $\Lambda > \Lambda_M$, it enters the strong-coupling regime, where the ground state becomes an entangled dressed state. 

Since $\rho \propto e^{-\beta H_S}$, the crossing changes the ordering of the thermal occupations and, accordingly, the construction of the passive state.
As a result, the total ergotropy takes two distinct analytical forms, given by
\begin{equation}\label{e}
\mathcal E(\rho) =
\begin{dcases}
\dfrac{2|\delta|}{\kappa Z} (\kappa -|\delta|) \sinh (\beta \kappa ),
& \Lambda\leq\Lambda_M ,
\\[6pt]
\dfrac{2 }{\kappa Z}
\begin{aligned}[t]
&\big((\kappa \omega-\delta ^2 )\sinh (\beta \kappa ) \\
&+\kappa (|\delta|-\omega ) \sinh (\beta \omega )\big)
\end{aligned},
& \Lambda>\Lambda_M .
\end{dcases}
\end{equation}
Equation~(\ref{e}) therefore separates the total ergotropy into two regimes distinguished by the ordering of the energy levels, and hence of the thermal occupations. This boundary will also play an important role in the decomposition of the total ergotropy into its incoherent and coherent parts, which we analyze later.

\subsection{Incoherent ergotropy}\label{Incoherent ergotropy}
We now evaluate the incoherent part of the ergotropy defined in Eq. \eqref{eq:E_i}, which is determined by the dephased state in the eigenbasis of $H_0$. From Eq.~\eqref{rho}, the corresponding dephased state is
\begin{multline}\label{rhodiag}
\pi=\frac{1}{Z}\mathrm{diag} \biggl(e^{-\beta \omega},\ \cosh (\beta \kappa )-\frac{\delta}{\kappa}\sinh (\beta \kappa ),\\
\cosh (\beta \kappa )+\frac{\delta}{\kappa}\sinh (\beta \kappa ),\ e^{\beta \omega }\biggr).
\end{multline}
The passive state of $\pi$ depends on the ordering of its diagonal populations.
In particular, the relevant changes occur when $\pi^{22}$ or $\pi^{33}$ becomes equal to $\pi^{44}$. Accordingly, we define two transition points.
The first one, $\Lambda_A$, is determined by $\pi^{22}=\pi^{44}$, namely
\begin{align}\label{A}
\kappa \cosh(\beta \kappa) - \delta \sinh(\beta \kappa) = \kappa e^{\beta \omega},
\end{align}
whereas the second one, $\Lambda_B$, follows from $\pi^{33}=\pi^{44}$, namely
\begin{align}\label{B}
\kappa \cosh(\beta \kappa) + \delta \sinh(\beta \kappa) = \kappa e^{\beta \omega}.
\end{align}
Note that $\kappa=\sqrt{4\Lambda^2+\delta^2}$ with $\delta = \omega_1-\omega_2$ and $\omega=\omega_1+\omega_2$. Hence, Eqs. \eqref{A} and \eqref{B} implicitly define the critical coupling strengths $\Lambda_A$ and $\Lambda_B$ that govern the population ordering in the dephased state $\pi$.

As shown in Fig.~\ref{98}(a), one finds $\Lambda_A < \Lambda_B$ for $\delta < 0$ and $\Lambda_A > \Lambda_B$ for $\delta > 0$. Figure~\ref{98}(b) indicates that this ordering is unchanged over the temperatures considered. This is because the sign of the detuning $\delta$ determines whether $\pi^{22}$ or $\pi^{33}$ is larger, which in turn determines the ordering of the critical couplings $\Lambda_A$ and $\Lambda_B$. For compactness, we therefore introduce
$\Lambda_{\min}=\min(\Lambda_A,\Lambda_B)$ and $\Lambda_{\max}=\max(\Lambda_A,\Lambda_B)$.
The incoherent ergotropy then takes the form
\begin{widetext}
\begin{equation}\label{ei}
\mathcal E_i(\rho) =
\begin{dcases}
0, & \Lambda \le \Lambda _{\min}, \\
\frac{(\omega-|\delta| ) }{\kappa Z}\left(-\kappa e^{\beta \omega }+|\delta| \sinh (\beta \kappa ) +\kappa \cosh (\beta \kappa )\right), & \Lambda _{\min}<\Lambda \le \Lambda _{\max}, \\
\frac{1}{\kappa Z}\bigl(-\kappa e^{\beta \omega } (\omega +|\delta| )+|\delta| ( \omega-3 |\delta| ) \sinh (\beta \kappa )+\kappa (\omega +|\delta|) \cosh (\beta \kappa )\bigr), & \Lambda >\Lambda _{\max}.
\end{dcases}
\end{equation}
\end{widetext}
Equation~\eqref{ei} shows that the incoherent ergotropy becomes nonzero only when the diagonal populations of the dephased state are reordered so that $\pi$ is no longer passive with respect to $H_0$.
Thus, the incoherent work extraction is governed by interaction-induced population reordering in the dephased state.

\subsection{Coherent ergotropy}\label{coherent ergotropy}

From Eqs.~\eqref{Delta}, \eqref{A}, and \eqref{B}, together with Fig.~\ref{98}(a), one finds that the characteristic coupling strength always satisfies $\Lambda_M < \Lambda_{\min}$. Combining this threshold with the critical coupling strengths $\Lambda_{\min}$ and $\Lambda_{\max}$ introduced above, the coherent ergotropy can be written in the form
\begin{widetext}
\begin{equation}\label{ec}
\mathcal E_c(\rho) = \begin{dcases}
\frac{2 |\delta| (\kappa-|\delta| )}{\kappa Z}\sinh (\beta \kappa ), & \Lambda \le\Lambda _M, \\
\frac{2 }{\kappa Z}\left(\left(\kappa \omega -\delta^2\right)\sinh (\beta \kappa ) +\kappa (|\delta| -\omega ) \sinh (\beta \omega )\right), & \Lambda _M<\Lambda \le \Lambda _{\min}, \\
\frac{1}{\kappa Z}\bigl(\kappa (\omega -|\delta| ) e^{-\beta \omega }+(2 \kappa \omega -|\delta| (|\delta| +\omega ))\sinh (\beta \kappa ) +\kappa (|\delta| -\omega ) \cosh (\beta \kappa )\bigr), & \Lambda _{\min}<\Lambda \le \Lambda _{\max}, \\
\frac{1}{\kappa Z}\bigl(\left(2 |\delta| e^{2 \beta \omega }-|\delta| +\omega \right)\kappa e^{-\beta \omega } +\left(\delta^2-|\delta| \omega +2 \kappa \omega \right)\sinh (\beta \kappa ) -\kappa (|\delta| +\omega ) \cosh (\beta \kappa )\bigr), & \Lambda >\Lambda _{\max}.
\end{dcases}
\end{equation}
\end{widetext}
\begingroup

The detailed derivation of Eqs.~\eqref{e}, \eqref{ei}, and \eqref{ec} and the corresponding passive state are provided in Appendix~\ref{app:passive_derivation}.
\endgroup
\hypersetup{linkcolor=blue,citecolor=blue}
Equation~\eqref{ec} shows that the coherent contribution is generally nonzero across these four intervals. In the interval $\Lambda_M < \Lambda \le \Lambda_{\min}$, one has $\mathcal E_c>0$ while $\mathcal E_i=0$, so that the ergotropy is purely coherent. This identifies a parameter window in which work extraction is enabled solely by coherence in the local energy basis.

\section{Resonant Charging Scheme}\label{balanced charging scheme}

Based on the analytical results obtained in Sec.~\ref{two TLS}, in this section we examine the magnitude and nature of the ergotropy in the resonant case where the two qubits have identical transition frequencies, denoted as $\omega_0 = \omega_1 = \omega_2$. We find that nonzero ergotropy is activated at $\Lambda_M$, defined in Eq.~\eqref{Delta}, whereas the incoherent contribution emerges only above the higher threshold $\Lambda_0>\Lambda_M$. Notably, within the intermediate coupling regime $\Lambda_M < \Lambda \le \Lambda_0$, the extractable work is entirely coherent. Once this threshold is exceeded, both incoherent and coherent ergotropy contribute. These two regimes are analyzed separately in Secs.~\ref{sec:ergotropy_coherent} and \ref{sec:ergotropy_mixed}, respectively.

\subsection{Ergotropy activation in the purely coherent regime}
\label{sec:ergotropy_coherent}

Under the resonant condition $\omega_0=\omega_{1}=\omega_{2}$, the analytical results obtained in Sec.~\ref{two TLS} simplify considerably. In particular, Eqs.~\eqref{A} and \eqref{B} reduce to the same equation, namely
\begin{align}
\cosh(\beta\kappa)=e^{\beta\omega}.
\end{align}
Therefore the two transition points $\Lambda_A$ and $\Lambda_B$ coincide. Accordingly, we introduce a single threshold
\begin{align}\label{lambda0}
\Lambda_{0}=\Lambda_{A}=\Lambda_{B}
=\frac{1}{2\beta}\cosh^{-1}\!\left(e^{\beta\omega}\right),
\end{align}
where we have applied $\delta=0$ and hence $\kappa=2\Lambda$. Moreover, the general ordering of the critical coupling strengths $\Lambda_M$, $\Lambda_A$, and $\Lambda_B$ established in Sec.~\ref{two TLS} implies $\Lambda_{M}<\Lambda_{0}$ also in the resonant case. Note that the threshold $\Lambda_M$ separating the weak and strong coupling regimes is given by Eq. \eqref{Delta}. 

\begin{figure}[t]
    \centering
    \includegraphics[width=\columnwidth]{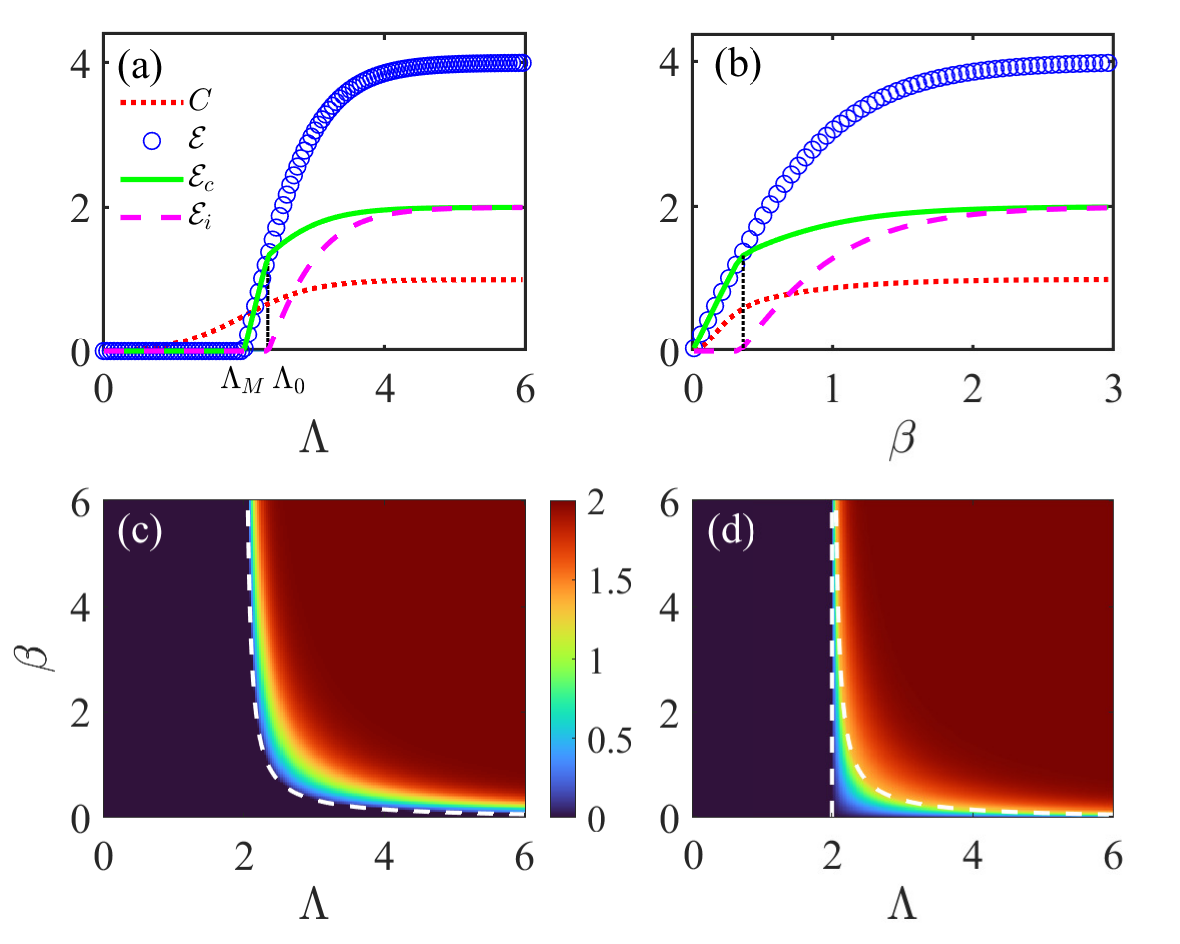}
    \caption{Ergotropy and coherence in the resonant case $\omega_1=\omega_2$. The total, coherent, and incoherent ergotropy, together with the quantum coherence $C$ defined in Eq.~\eqref{C}, are shown as functions of (a) the coupling strength $\Lambda$ at $\beta=1$, and (b) the inverse temperature $\beta$ at $\Lambda=3$. Panels (c) and (d) display the incoherent and coherent ergotropy in terms of the coupling strength $\Lambda$ and the inverse temperature $\beta$, respectively. The white dashed line in (c) denotes the critical threshold $\Lambda_0$, while those in (d) mark the boundaries $\Lambda_M$ and $\Lambda_0$. The individual qubit frequencies are fixed at $\omega_1=\omega_2=1$, so that $\omega=2$.}\label{14a15a}
 \end{figure}

We first consider the regime $\Lambda\leq\Lambda_{M}$. In this interval, the ergotropy remains zero,
\begin{align}\label{ec1w0}
\mathcal{E}=\mathcal{E}_{i}=\mathcal{E}_{c}=0.
\end{align}
\begingroup

As shown in Fig.~\ref{14a15a}(a), the thermal state
already contains finite off-diagonal coherence in the local product
basis within the single-excitation subspace. At resonance,
$|01\rangle$ and $|10\rangle$ are degenerate with respect to $H_0$,
and the XX interaction commutes with the bare Hamiltonian,
$[H_0,H_I]=0$. The interaction therefore rotates states only within
this degenerate $H_0$ subspace, so the resulting basis coherence
carries no bare-energy cost. Moreover, for $\Lambda\leq\Lambda_M$,
the ordering of the Gibbs eigenvalues remains compatible with the
passive ordering relative to $H_0$. The charged state therefore
remains passive and its ergotropy vanishes, despite the presence of
off-diagonal coherence in the chosen local basis. Thus, in the
resonant case, such coherence alone does not activate extractable
work; non-passivity requires the additional spectral reordering that
occurs at $\Lambda_M$.
\endgroup
\hypersetup{linkcolor=blue,citecolor=blue}

For $\Lambda_{M}<\Lambda\leq\Lambda_{0}$, the charged energy can be extracted, and the ergotropy takes the form
\begin{align}\label{ec2w0}
\mathcal{E}=\mathcal{E}_{c}
=\frac{2\omega}{Z}\left(\sinh(\beta\kappa)-\sinh(\beta\omega)\right),
\end{align}
while the incoherent ergotropy remains zero, namely
\begin{align}\label{e2w0}
\mathcal{E}_{i}=0.
\end{align}
Although the interaction-induced coherence is sufficient to activate a finite ergotropy, the dephased state remains passive, so that no incoherent contribution is yet available. This identifies $\Lambda_{0}$ as the threshold at which incoherent ergotropy starts to emerge, as discussed in the next subsection.

In the zero-temperature limit $\beta \to \infty$, the unified boundary approaches $\Lambda_0 \to \omega/2$. Evaluating the asymptotic form of the hyperbolic functions in this regime, one finds that the ergotropy approaches the temperature-independent constant
\begin{align}
\lim_{\beta\to\infty}\mathcal{E}=\frac{\omega}{3}.
\end{align}
By contrast, in the high-temperature limit $\beta \to 0$, a Taylor expansion yields
\begin{align}
\mathcal{E}=\mathcal{E}_c \simeq \frac{1}{2}\omega\beta(\kappa-\omega),
\end{align}
which vanishes linearly with $\beta$ as thermal fluctuations become dominant. Therefore, within this purely coherent regime, the low-temperature limit leads to a finite saturation value of the ergotropy, whereas increasing temperature progressively suppresses the extractable work.

We now analyze the limiting behaviors of the ergotropy in the purely coherent regime. At $\Lambda=\Lambda_{0}$, Eq.~\eqref{ec2w0} gives
\begin{align}\label{lambda0ec2}
\mathcal{E}_{c}=
\frac{\omega \left( \sqrt{e^{3\beta\omega} - 1} - \sinh(\beta\omega) \right)}{e^{3\beta\omega} + \cosh(\beta\omega)}.
\end{align}
Along the boundary $\Lambda=\Lambda_{0}$, one has
\begin{equation}
\kappa=\frac{1}{\beta}\cosh^{-1}(e^{\beta\omega}) \to \omega
\qquad (\beta\to\infty),
\end{equation}
and the ergotropy approaches
\begin{align}
  \lim_{\beta \to \infty}\mathcal{E}
  =\lim_{\beta \to \infty}\mathcal{E}_{c}
  =\frac{\omega}{3}.
\end{align}
In the high-temperature limit $\beta \to 0$, one finds $\kappa \to \sqrt{2\omega/\beta}$, so that
\begin{align}
\mathcal{E}_{c}\simeq \frac{\omega}{2}\sqrt{2\beta \omega}\to 0.
\end{align}
These limiting behaviors are consistent with the numerical results shown in Fig.~\ref{14a15a}(b).

The fact that coherent ergotropy is activated before its incoherent counterpart reflects the threshold hierarchy discussed above. In the region $\Lambda_{M}<\Lambda\leq\Lambda_{0}$, the coupling strength is already sufficient to generate coherence that makes the full state non-passive, but it is still insufficient to modify the population distribution of the dephased state to the same extent. As a result, only the coherent contribution makes the battery active, while the incoherent contribution remains zero. This is why the coherent contribution appears first, whereas the incoherent contribution emerges only at a larger coupling.

\subsection{Ergotropy activation in the mixed regime}
\label{sec:ergotropy_mixed}

When $\Lambda > \Lambda_{0}$, both coherent and incoherent ergotropy contribute to the total extractable work. In this regime, the coherent and incoherent parts are given by
\begin{align}\label{ec3w0}
\mathcal{E}_{c}
=\frac{\omega}{Z}\left(2 \sinh (\beta \kappa )-\cosh (\beta \kappa )+e^{-\beta \omega }\right),
\end{align}
\begin{align}\label{ei3w0}
\mathcal{E}_{i}
=\frac{\omega}{Z}\left(\cosh (\beta \kappa )-e^{\beta \omega }\right).
\end{align}
The total ergotropy is therefore
\begin{equation}
\label{ec2w0_equal}
\mathcal{E} = \mathcal{E}_c + \mathcal{E}_i
= \frac{2\omega}{Z}\left(\sinh(\beta\kappa)-\sinh(\beta\omega)\right),
\end{equation}
which has the same form as in the interval $\Lambda_{M}<\Lambda\leq\Lambda_{0}$ given by Eq.~\eqref{ec2w0}. Thus, the total ergotropy varies continuously across $\Lambda_{0}$, while its decomposition changes because the incoherent contribution becomes nonzero only for $\Lambda>\Lambda_{0}$. 

For fixed $\Lambda>\Lambda_{0}$, equivalently $\kappa>\cosh^{-1}(e^{\beta\omega})/\beta$, the asymptotic behavior is qualitatively different. In the zero-temperature limit $\beta \to \infty$, the total ergotropy approaches
\begin{align}
  \lim_{\beta \to \infty}\mathcal{E} = \omega,
\end{align}
with equal coherent and incoherent contributions,
\begin{align}
  \lim_{\beta \to \infty}\mathcal{E}_{i}
  =\lim_{\beta \to \infty}\mathcal{E}_{c}
  =\frac{\omega}{2},
\end{align}
as shown in Fig.~\ref{14a15a}(c) and (d). In the high-temperature limit $\beta \to 0$, for fixed $\Lambda>\Lambda_{0}$ we obtain
\begin{align}
\mathcal{E}_{c}\simeq \mathcal{E}\simeq \frac{\omega(2\kappa-\omega)}{4}\beta,
\end{align}
which vanishes as $\beta\to 0$. The above result shows that, for fixed $\Lambda>\Lambda_{0}$, the low-temperature limit drives the resonant charging scheme into a fully activated regime, where the extractable work reaches its maximal value $\mathcal{E}=\omega$. Physically, as $\beta\to\infty$, the dissipative dynamics prepares a state that is increasingly close to the ground state of the interacting Hamiltonian, which is strongly non-passive with respect to the bare Hamiltonian $H_{0}$ after the interaction is switched off. Moreover, the relation $\mathcal{E}_{c}=\mathcal{E}_{i}=\omega/2$ indicates an exact equipartition between the coherent and incoherent parts of work extraction. This shows that, in the resonant mixed regime at low temperature, quantum coherence and population redistribution become equally important resources for ergotropy generation. In this sense, the threshold $\Lambda_{0}$ does not merely signal the appearance of incoherent ergotropy, but also marks the regime from which the battery can evolve toward a maximally active state with balanced coherent and incoherent contributions.

We next consider the strong-coupling limit $\Lambda\to\infty$ at fixed $\beta$. In this case,
\begin{align}
\mathcal{E} \simeq \frac{\omega\left[1-2\sinh(\beta \omega)e^{-\beta \kappa}\right]}
{1+2\cosh(\beta \omega)e^{-\beta \kappa}},
\end{align}
so that $\mathcal{E}\to\omega$ as $\kappa\to\infty$. Likewise,
\begin{align}
\mathcal{E}_{c}
\simeq \frac{\omega(e^{\beta \kappa}+2e^{-\beta \omega})}
{2e^{\beta \kappa}+4\cosh(\beta \omega)}
\to \frac{\omega}{2},
\end{align}
and
\begin{align}
\mathcal{E}_{i}
\simeq \frac{\omega(e^{\beta \kappa}-2e^{-\beta \omega})}
{2e^{\beta \kappa}+4\cosh(\beta \omega)}
\to \frac{\omega}{2}.
\end{align}
Therefore the equipartition is not restricted to the low-temperature limit; it also emerges in the strong-coupling limit at fixed $\beta$. Table~\ref{twoomega0} summarizes the main results in this section, which clearly shows the different regimes of ergotropy generation and their dependence on the coupling strength and temperature.

\begin{table}[t]
  \centering
  \caption{Summary of the ergotropy under the resonant condition $\omega_0=\omega_1=\omega_2$, with $\omega=\omega_1+\omega_2=2\omega_0$. The threshold $\Lambda_{0}$ given by Eq. \eqref{lambda0} separates the purely coherent regime from the mixed regime, while $\Lambda_{M}$ given by Eq. \eqref{Delta} determines the activation of nonzero total ergotropy.}
  \begin{tabular}{c|c|ccc}
  \hline\hline
    $\Lambda$ & $\beta$ & $\mathcal{E}$ & $\mathcal{E}_c$ & $\mathcal{E}_i$ \\ \hline
    $0\le \Lambda \le \Lambda_M$ & Finite $\beta$ & 0 & 0 & $0$ \\\hline
    \multirow{2}{*}{$\Lambda_M< \Lambda < \Lambda_0$} & Finite $\beta$ &  Eq. (\ref{ec2w0}) & Eq. (\ref{ec2w0}) & 0 \\ 
     & $ \beta\to \infty$ & $\omega/3$ & $\omega/3$ & $0$ \\ 
     \hline
    \multirow{2}{*}{$ \Lambda =\Lambda_0$} & Finite $\beta$ &  Eq. (\ref{lambda0ec2}) & Eq. (\ref{lambda0ec2}) & 0 \\ 
     & $ \beta\to \infty$ & $\omega/3$ & $\omega/3$ & $0$ \\ 
    \hline
    \multirow{2}{*}{$\Lambda > \Lambda_0$} & Finite $\beta$ & Eq. (\ref{ec2w0_equal}) & Eq. (\ref{ec3w0})& Eq. (\ref{ei3w0}) \\
     & $ \beta\to \infty$ &$\omega$ & $\omega/2$ & $\omega/2$\\ 
     \hline
    \multirow{2}{*}{$\Lambda \to \infty$} & Finite $\beta$ & $\omega$ & $\omega/2$& $\omega/2$ \\
   & $ \beta\to \infty$ &$\omega$ & $\omega/2$ & $\omega/2$\\ 
    \hline\hline
  \end{tabular}
  \label{twoomega0}
\end{table}

\section{Detuned Charging Scheme}\label{detuning Charging scheme}

In this section, we consider the detuned charging scheme, where the two qubits have different frequencies. Without loss of generality, we take $\omega_2>\omega_1$ (equivalently, $\delta=\omega_1-\omega_2<0$). To clarify the physical role of frequency mismatch, we analyze the charging performance in three regimes separated by the coupling strength: the weak-coupling regime (Sec.~\ref{E*M}), the critical coupling strength with $\Lambda=\Lambda_M$ (Sec.~\ref{nL*M}), and the strong-coupling regime (Sec.~\ref{nbL*M}).
As will be shown, detuning qualitatively modifies the charging behavior in three respects: it activates purely coherent ergotropy in the weak-coupling regime, yields a finite zero-temperature coherent contribution at the critical point, and changes the partition of the total ergotropy between coherent and incoherent parts.

\begin{figure}[t]
    \centering
    \includegraphics[width=\columnwidth]{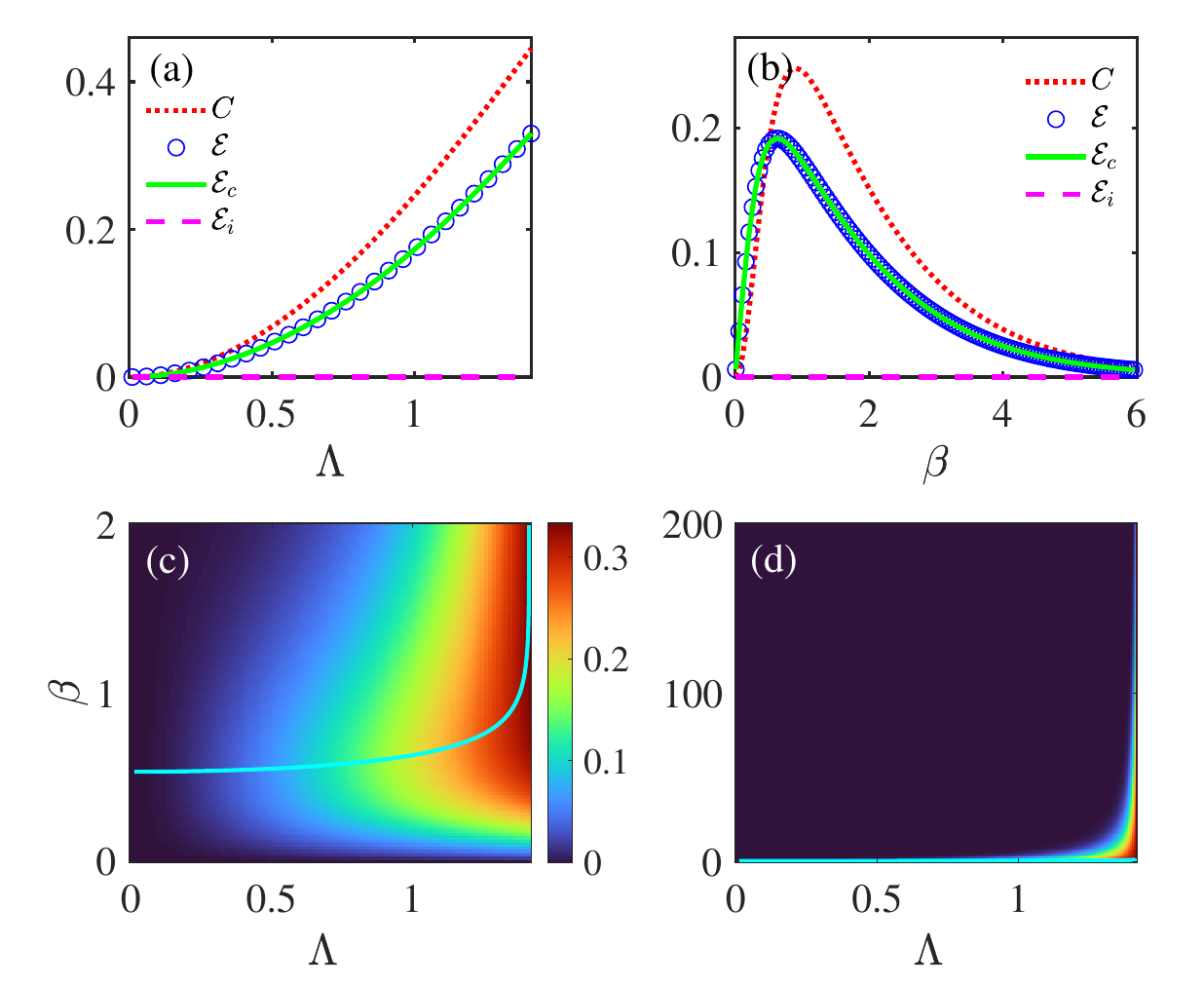}
     \caption{Ergotropy and coherence in the detuned charging scheme. The total, coherent, and incoherent ergotropy, together with the quantum coherence $C$ defined in Eq.~\eqref{C}, are shown as functions of (a) the coupling strength $\Lambda$ at $\beta=1$ and (b) the inverse temperature $\beta$ at $\Lambda=1$. Panels (c) and (d) display the ergotropy in the low-temperature regime and in the zero-temperature limit $\beta\to\infty$, respectively. The sky-blue line in (c) marks the maximum coherent ergotropy explicitly. The qubit frequencies are set to $\omega_1=1$ and $\omega_2=2$.}\label{713a}
 \end{figure}

\subsection{Ergotropy activation in the weak-coupling regime}\label{E*M}

In the weak-coupling regime ($\Lambda<\Lambda_M$) with $\omega_1<\omega_2$, ergotropy can already be generated, in sharp contrast to the resonant case. More specifically, the extractable work is purely coherent in this regime, and the total and coherent ergotropy coincide, given by
\begin{align}\label{ec1}
\mathcal{E}_{c}=\mathcal{E}
=-\frac{2\delta(\delta+\kappa)}{\kappa Z}\sinh(\beta\kappa),
\end{align}
where $Z=2\left(\cosh(\beta\kappa)+\cosh(\beta\omega)\right)$ and $\kappa=\sqrt{4\Lambda^2+\delta^2}$.
In the weak-coupling regime with $0<\Lambda<\Lambda_M$, one has
\begin{equation}
|\delta|<\kappa<\omega.
\end{equation}
\begingroup

Thus, the eigenenergies of $H_S$ retain the ordering $-\omega<-\kappa<\kappa<\omega$, and no spectral crossing occurs. At resonance, the single-excitation configurations $|01\rangle$ and $|10\rangle$ are degenerate with respect to $H_0$. The XX interaction therefore mixes them only within an equal-energy subspace, so this basis rotation carries no bare-energy cost. Together with the unchanged Gibbs ordering, this ensures that the resonant weak-coupling state remains passive with respect to $H_0$.

Detuning breaks the single-excitation degeneracy and the exchange symmetry, causing the interaction to coherently mix states with different bare energies. At finite temperature, the resulting mismatch between the Gibbs eigenbasis of $H_S$ and the energy eigenbasis of $H_0$ makes the state non-passive even though the dressed-level ordering remains unchanged. Since the dephased state $\pi$ remains passive in this regime, $\mathcal E_i=0$ and the activated ergotropy is entirely coherent.
\endgroup

The extrema of $\mathcal E_c$ with respect to temperature are determined by $\partial\mathcal E_c/\partial\beta=0$, which gives
\begin{align}
\frac{\kappa}{\tanh(\beta\kappa)}
=
\frac{\kappa\sinh(\beta\kappa)+\omega\sinh(\beta\omega)}
{\cosh(\beta\kappa)+\cosh(\beta\omega)}.
\end{align}
This condition determines the maximum coherent ergotropy in the $(\Lambda,\beta)$ parameter space, shown by the sky-blue line in Fig.~\ref{713a}(c). As illustrated in Fig.~\ref{713a}(b), the coherent ergotropy is not a monotonic function of the coherence measure $C$. The reason is that $\mathcal E_c$ depends not only on the magnitude of coherence, but also on the population distribution in the energy basis. Therefore, thermal fluctuations may enhance $C$ while weakening the population required for work extraction.

The zero-temperature behavior follows directly from the inequality $\kappa<\omega$. In the limit $\beta\to\infty$, Eq.~\eqref{ec1} reduces to
\begin{align}\label{ec1wuqiong}
\mathcal{E}_{c}
\simeq
-\frac{\delta(\delta+\kappa)}{\kappa}e^{\beta(\kappa-\omega)}
\to 0.
\end{align}
\begingroup

This exponential suppression is consistent with Fig.~\ref{713a}(d). Physically, in the weak-coupling regime, the thermal state approaches the ground state of $H_S$ as $\beta\to\infty$, and this ground state coincides with the bare product state $\rho=|11\rangle\langle 11|$. Since it carries neither local excitation nor useful coherence, the extractable work vanishes in the zero-temperature limit.
\endgroup
\hypersetup{linkcolor=blue,citecolor=blue}

In the opposite high-temperature limit $\beta\to0$, Eq.~\eqref{ec1} gives
\begin{align}\label{ec10}
\mathcal{E}_{c}
\simeq
-\frac{\delta(\delta+\kappa)}{2}\beta
+\mathcal{O}(\beta^3)
\to 0,
\end{align}
in agreement with the behavior shown in Fig.~\ref{713a}(b). Moreover, in the vanishing-interaction limit $\Lambda\to0$, where $\kappa\to|\delta|$, Eq.~\eqref{ec1} yields
\begin{align}
\lim_{\Lambda\to0}\mathcal E_c=0,
\end{align}
consistent with Fig.~\ref{713a}(a).

\subsection{Ergotropy activation at the critical coupling strength}\label{nL*M}

At the critical point $\Lambda=\Lambda_M$ (equivalently, $\kappa=\omega$), the coherent ergotropy has the exact expression
\begin{equation}\label{LAMBDAM}
\mathcal{E}_{c}=\frac{(\omega_2-\omega_1)\omega_1}{\omega_1+\omega_2}
\tanh\!\left((\omega_1+\omega_2)\beta\right).
\end{equation}
This result immediately yields two limiting behaviors. In the zero-temperature limit $\beta\to\infty$, one obtains
\begin{equation}\label{LAMBDAMINFTY}
\lim_{\beta\to\infty}\mathcal{E}_{c}
=
\frac{(\omega_2-\omega_1)\omega_1}{\omega_1+\omega_2},
\end{equation}
showing that the coherent ergotropy remains finite at the critical point. In the opposite high-temperature limit $\beta\to0$, it vanishes,
\begin{equation}
\lim_{\beta\to0}\mathcal{E}_{c}=0.
\end{equation}
These limits, consistent with Fig.~\ref{713a}(c), highlight the special role of the critical coupling. In the weak-coupling regime, the zero-temperature limit yields $\mathcal E_c\to0$, as shown in Figs.~\ref{713a}(b) and \ref{713a}(d). At $\Lambda=\Lambda_M$, a finite coherent contribution survives even as $\beta\to\infty$. This contrast indicates a nonuniform zero-temperature behavior at the critical point: coherent work extraction is absent for $\Lambda<\Lambda_M$ but remains finite exactly at $\Lambda=\Lambda_M$. By contrast, the high-temperature limit $\beta\to0$ suppresses coherent ergotropy in all cases.

\begin{figure}[t]
    \centering
    \includegraphics[width=\columnwidth]{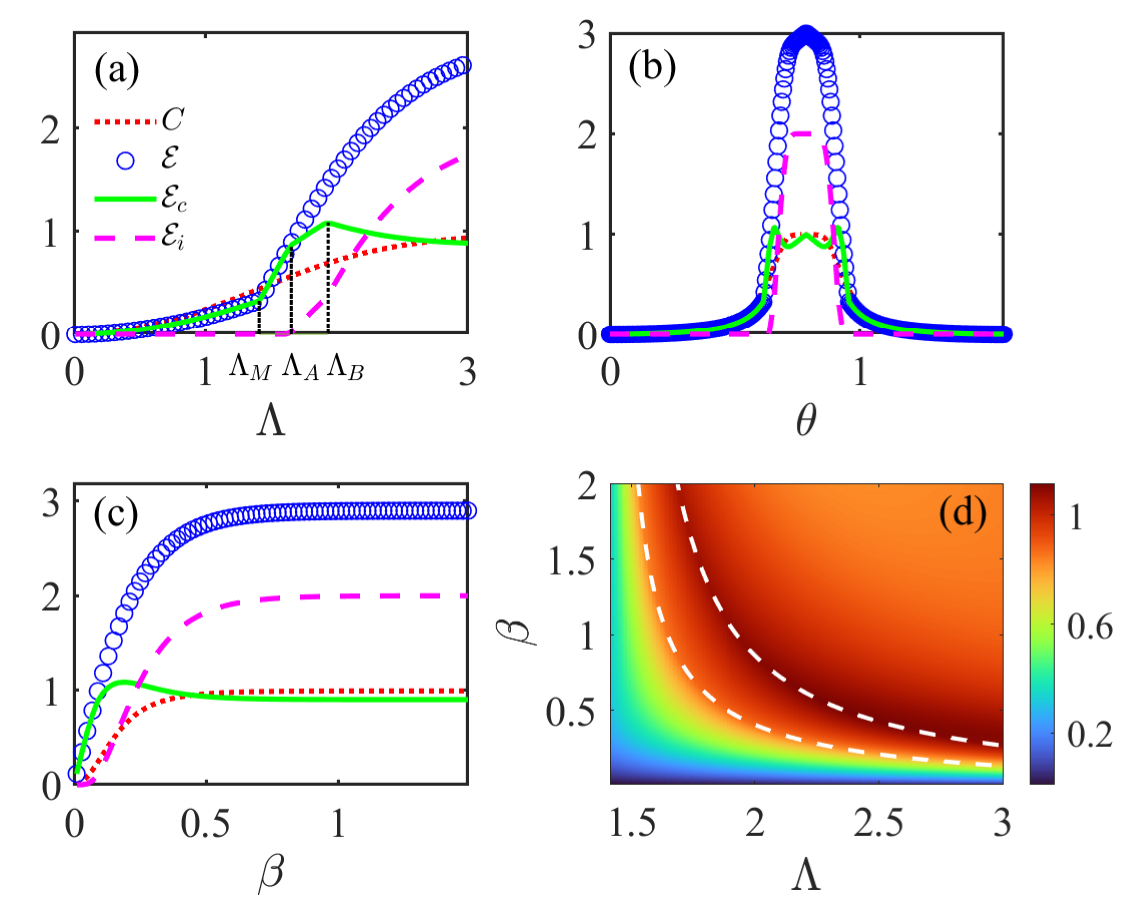}
     \caption{Ergotropy and coherence in the detuned charging scheme. The total, coherent, and incoherent ergotropy, together with the quantum coherence $C$ defined in Eq.~\eqref{C}, are shown as functions of (a) the coupling strength $\Lambda$ and (b) the mixing angle $\theta$, both at $\beta=1$. Panel (c) displays the total, incoherent, and coherent ergotropy as functions of the inverse temperature $\beta$ at $\Lambda=5$. Panel (d) shows the total ergotropy as a function of both $\Lambda$ and $\beta$. The white dashed lines in (d) mark the critical coupling strengths $\Lambda_A$ and $\Lambda_B$ given by Eqs.~\eqref{A} and \eqref{B}, respectively. The qubit frequencies are set to $\omega_1=1$ and $\omega_2=2$.}\label{2xinbc}
\end{figure}

\subsection{Ergotropy activation in the strong-coupling regime}\label{nbL*M}

We now turn to the strong-coupling regime, where coherent and incoherent contributions coexist. As shown in Fig.~\ref{2xinbc}(a) and (c), within part of this regime the quantum coherence and the coherent ergotropy vary nonmonotonically with the coupling strength. This behavior is qualitatively consistent with the general observation that coherence is not, by itself, a monotonic indicator of coherent ergotropy \cite{PhysRevLett.125.180603}.

Figure~\ref{2xinbc}(b) further shows how ergotropy and coherence vary with the mixing angle $\theta$, where the maximum is achieved at $\theta=\pi/4$. This behavior suggests that the resonant case is the most favorable for coherent work extraction. As $\theta$ moves away from $\pi/4$, the two-qubit model becomes less symmetric, and the cooperative interplay between coherence generation and energy redistribution is weakened. The $\theta$ dependence therefore shows that, besides the interaction strength and detuning, the efficiency of the charging process is also shaped by the mixing structure of the two local modes.

Next we consider the zero-temperature limit $\beta\to\infty$, by which the critical coupling strengths $\Lambda_A$ and $\Lambda_B$ defined in Eqs. \eqref{A} and \eqref{B} reduce to
\begin{align}
(\kappa-\delta)e^{\beta\kappa}=2\kappa e^{\beta\omega},
\qquad
(\kappa+\delta)e^{\beta\kappa}=2\kappa e^{\beta\omega},
\end{align}
respectively. These relations can only be satisfied asymptotically when $\kappa\simeq\omega$, indicating that $\Lambda_M$, $\Lambda_A$, and $\Lambda_B$ approach one another in the low-temperature limit. Throughout the strong-coupling region considered here, one has $\kappa>\omega$. The ergotropy components therefore approach
\begin{align}\label{wcin}
\lim_{\beta\to\infty}\mathcal{E}_{c}
\simeq
\frac{(\delta+\kappa)(\delta+\omega)}{2\kappa}
=
\omega_1\left(1+\frac{\delta}{\kappa}\right),
\end{align}
\begin{align}\label{wiin}
\lim_{\beta\to\infty}\mathcal{E}_{i}
\simeq
-\frac{\delta(\delta+\omega_1)}{\kappa}+\omega_2,
\end{align}
\begin{align}\label{win}
\lim_{\beta\to\infty}\mathcal{E}
\simeq
\omega-\frac{\delta^2}{\kappa},
\end{align}
as shown in Fig.~\ref{2xinbc}(c). In the further limit $\Lambda\to\infty$, one has $\delta/\kappa\simeq\delta/(2\Lambda)\to0$, so that
\begin{align}
\mathcal{E}_{c}\to\omega_1,\qquad
\mathcal{E}_{i}\to\omega_2,\qquad
\mathcal{E}\to\omega.
\end{align}
Thus, unlike the resonant case, the detuned scheme approaches the maximal total ergotropy only asymptotically as $\Lambda\to\infty$.

In the high-temperature limit $\beta\to0$, the defining conditions for $\Lambda_A$ and $\Lambda_B$ reduce to $\omega=-\delta$ and $\omega=\delta$, respectively, which admit no solution. Hence the incoherent contribution of ergotropy disappears so that $\mathcal{E}_{c}\simeq\mathcal{E}$. Expanding Eq.~\eqref{e} for small $\beta$, we obtain
\begin{align}
\lim_{\beta\to0}\mathcal{E}
\simeq
\frac{1}{2}\left[(\kappa-\omega)\omega-(\delta+\omega)\delta\right]\beta
\to0,
\end{align}
consistent with Figs.~\ref{2xinbc}(c) and \ref{2xinbc}(d).

For arbitrary finite temperature with $\beta\neq0$, the asymptotic limit $\Lambda\to\infty$ gives
\begin{align}
\lim_{\Lambda\to\infty}\mathcal{E}_{c}
\simeq
\frac{\omega+\delta}{2}
=
\omega_1,
\end{align}
\begin{align}
\lim_{\Lambda\to\infty}\mathcal{E}_{i}
\simeq
\frac{\omega-\delta}{2}
=
\omega_2,
\end{align}
\begin{align}
\lim_{\Lambda\to\infty}\mathcal{E}
\simeq
\omega-\frac{\delta^2}{\kappa}
\simeq
\omega,
\end{align}
as illustrated in Fig.~\ref{2xinbc}(a). These asymptotic limits show that, in contrast to the resonant scheme where the coherent and incoherent parts become equally shared, the detuned configuration partitions the extractable work according to the two qubit frequencies, with $\mathcal E_c\to\omega_1$ and $\mathcal E_i\to\omega_2$. Such a partition can be understood as a direct consequence of the broken exchange symmetry induced by detuning. In the resonant case, the two subsystems are energetically equivalent, so the strong-coupling limit leads to a symmetric division between coherent and incoherent contributions. By contrast, when $\omega_1\neq\omega_2$, the two local energy scales are no longer identical, and the asymptotic work content carried by the coherent and incoherent contributions becomes locked to these distinct frequencies. As a result, the coherent part asymptotically approaches the lower local scale $\omega_1$, while the incoherent part approaches the higher one $\omega_2$, and their sum recovers the maximal total ergotropy $\omega_1+\omega_2$. The main results for the detuned case are summarized in Table~\ref{omega12}.

\begin{table}[t]
  \centering
  \caption{Summary of ergotropy in the detuned case $\omega_1<\omega_2$. The critical coupling strength $\Lambda_M$ separates the purely coherent weak-coupling regime from the regime where coherent and incoherent contributions coexist.}
  \begin{tabular}{c|c|ccc}
  \hline\hline
    $\Lambda$ & $\beta$ & $\mathcal{E}$ & $\mathcal{E}_c$ & $\mathcal{E}_i$ \\ \hline
    $\Lambda=0$ & Finite $\beta$ & $0$ & $0$ & $0$ \\
    \hline
    \multirow{2}{*}{$0<\Lambda<\Lambda_M$}
    & Finite $\beta$ & Eq.~(\ref{ec1}) & Eq.~(\ref{ec1}) & $0$ \\
    & $\beta\to\infty$ & $0$ & $0$ & $0$ \\
    \hline
    \multirow{2}{*}{$\Lambda=\Lambda_M$}
    & Finite $\beta$ & Eq.~(\ref{LAMBDAM}) & Eq.~(\ref{LAMBDAM}) & $0$ \\
    & $\beta\to\infty$ & Eq.~(\ref{LAMBDAMINFTY}) & Eq.~(\ref{LAMBDAMINFTY}) & $0$ \\
    \hline
    \multirow{2}{*}{$\Lambda>\Lambda_M$}
    & Finite $\beta$ & Eq.~(\ref{e}) & Eq.~(\ref{ec}) & Eq.~(\ref{ei}) \\
    & $\beta\to\infty$ & Eq.~(\ref{win}) & Eq.~(\ref{wcin}) & Eq.~(\ref{wiin}) \\
    \hline
    \multirow{2}{*}{$\Lambda\to\infty$}
    & Finite $\beta$ & $\omega$ & $\omega_1$ & $\omega_2$ \\
    & $\beta\to\infty$ & $\omega$ & $\omega_1$ & $\omega_2$ \\
    \hline\hline
  \end{tabular}
  \label{omega12}
\end{table}

\section{Ergotropy Activation in the \texorpdfstring{$N$}{N}-TLS XX spin chain}

\label{XX spin}

Having established the ergotropy properties of the two-qubit battery, we now extend the analysis to the $N$-TLS XX spin chain with periodic boundary conditions. We first examine the three-TLS case in Sec.~\ref{three}, and then generalize the discussion to arbitrary $N$ in Sec.~\ref{N TLS}, showing that increasing system size promotes interaction-induced level crossings and, consequently, ergotropy generation.

\subsection{XX spin chain with three TLSs}\label{three}

\begin{figure}[t]
    \centering
\includegraphics[width=3.3in]{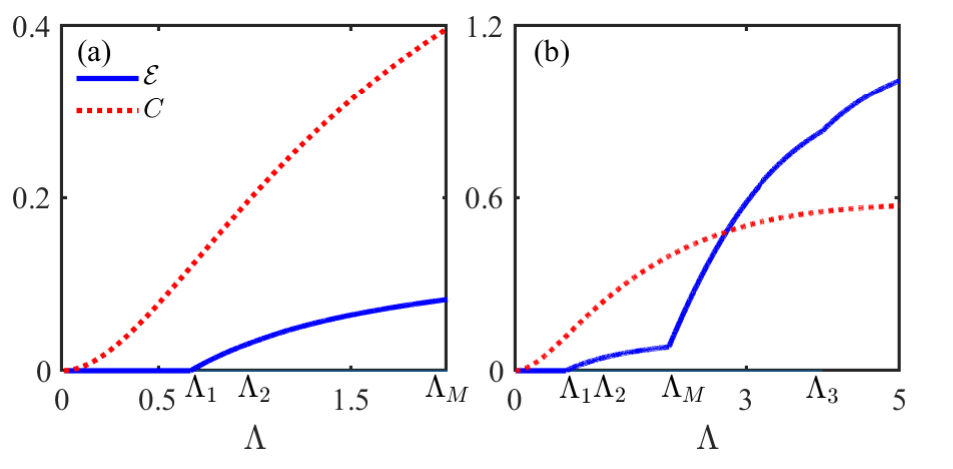}
     \caption{
     The total ergotropy and quantum coherence in the three-TLS XX spin chain as functions of the coupling strength $\Lambda$ at $\beta=1$. Panel (a) covers the interval $0\leq \Lambda \leq \Lambda_M$, while panel (b) covers $\Lambda_M \leq \Lambda \leq 5$. The horizontal-axis markers $\Lambda_i$ denote the successive energy-level crossing points, with $\Lambda_1$ corresponding to the first crossing. The threshold $\Lambda_M$ marks the crossing between the ground state and the first excited state. The parameters are set to $N=3$ and $\omega_0=1$.}
  \label{6666}
\end{figure}

We now consider the case $N=3$ under periodic boundary conditions, and assume identical onsite energies $\omega_0$ for all battery spins. In the eigenbasis of $H_S$, the Hamiltonian $\widetilde{H}_S$ is diagonal, with energy levels
\begin{subequations}
\begin{align}
E_1&=-3\omega_0,\\
E_2&=E_3=-\Lambda-\omega_0,\\
E_4&=2\Lambda-\omega_0,\\
E_5&=E_6=-\Lambda+\omega_0,\\
E_7&=2\Lambda+\omega_0,\\
E_8&=3\omega_0.
\end{align}
\end{subequations}
From this spectrum, the crossing between the ground state $E_1$ and the first excited level $E_2$ occurs at
\begin{align}\label{N3M}
\Lambda_M = 2\omega_0.
\end{align}
For $\Lambda<\Lambda_M$, the ground state remains nondegenerate and separated, namely the state $|111\rangle$. In the low-temperature limit $\beta\to\infty$, the thermal state therefore approaches the ground state of $H_S$, which is passive with respect to $H_0$, so the ergotropy vanishes. By contrast, for $\Lambda>\Lambda_M$, the lowest-energy manifold becomes twofold degenerate. The corresponding low-temperature thermal state is then an equal mixture, which is non-passive with respect to $H_0$ and hence yields nonzero ergotropy.

Notably, the ground-state crossing is not the first level crossing in the spectrum when increasing the coupling strength $\Lambda$. Instead, the first crossing occurs between the excited levels $E_4$ and $E_5$, which gives
\begin{align}
\Lambda_1 = \frac{2}{3}\omega_0,
\end{align}
so that $\Lambda_1<\Lambda_M$, as shown in Fig.~\ref{6666}. This example already shows that for $N>2$, the first interaction-induced level crossing need not involve the ground state and first excited state, but can instead arise within the excited-state spectrum. As discussed below, such a pre-$\Lambda_M$ crossing can already be accompanied by the onset of ergotropy generation.

\begingroup

\subsection{XX spin chain with \texorpdfstring{$N$}{N} TLSs}
\label{N TLS}

We now determine analytically the two interaction scales that organize the many-body ergotropy activation. They have distinct definitions and thermodynamic roles. The first, $\Lambda_1$, is the earliest positive level crossing in the full many-body spectrum and therefore sets the boundary for finite-temperature ergotropy activation. The second, $\Lambda_M$, is the crossing at which the fully polarized ground state is replaced by the lowest one-excitation state. For $N>2$, $\Lambda_1$ marks the first rearrangement within the excited spectrum, whereas $\Lambda_M$ marks a reconstruction of the ground state.

The distinction follows directly from excitation-number conservation. In the subspace $\mathcal H_n$ containing $n$ excitations, the bare Hamiltonian is proportional to the identity, $H_0|_{\mathcal H_n}=(-N+2n)\omega_0 I_{\mathcal H_n}$, whereas $H_I|_{\mathcal H_n}=\Lambda V_n$. Here $V_n$ is the dimensionless interaction matrix in $\mathcal H_n$, obtained by factoring $\Lambda$ out of $H_I$. If $\mu_{n,\alpha}$ is an eigenvalue of $V_n$, every many-body energy level is therefore exactly affine in the interaction strength,
\begin{equation}
E_{n,\alpha}(\Lambda)=(-N+2n)\omega_0+\mu_{n,\alpha}\Lambda .
\label{eq:N_affine}
\end{equation}
Consequently, all positive crossing points can be obtained from intersections of straight spectral lines, without scanning $\Lambda$ on a numerical grid.

For a homogeneous periodic chain, the complete analytical results are
\begin{equation}
\Lambda_M=
\begin{cases}
\omega_0, & N\ \text{even},\\[1mm]
\omega_0\sec\left(\dfrac \pi N\right), & N\ \text{odd},
\end{cases}
\label{eq:N_LM}
\end{equation}
and
\begin{equation}
\Lambda_1=
\begin{cases}
\omega_0\tan\!\left(\dfrac{\pi}{2N}\right), & N\ \text{even},\\[2mm]
\omega_0\tan\!\left(\dfrac{\pi}{2N}\right)
\sec\!\left(\dfrac{\pi}{2N}\right), & N\ \text{odd}.
\end{cases}
\label{eq:N_L1}
\end{equation}
The derivation, including the parity-dependent Jordan--Wigner momentum grids and a proof that no crossing between non-neighboring excitation subspaces can occur at a smaller positive coupling, is given in Appendix~\ref{app:xx_thresholds}. As a consistency check, Eqs.~\eqref{eq:N_LM} and \eqref{eq:N_L1} reproduce the three-spin values $\Lambda_M=2\omega_0$ and $\Lambda_1=2\omega_0/3$ obtained above.

\begin{figure}[t]
    \centering
    \includegraphics[width=3.3in]{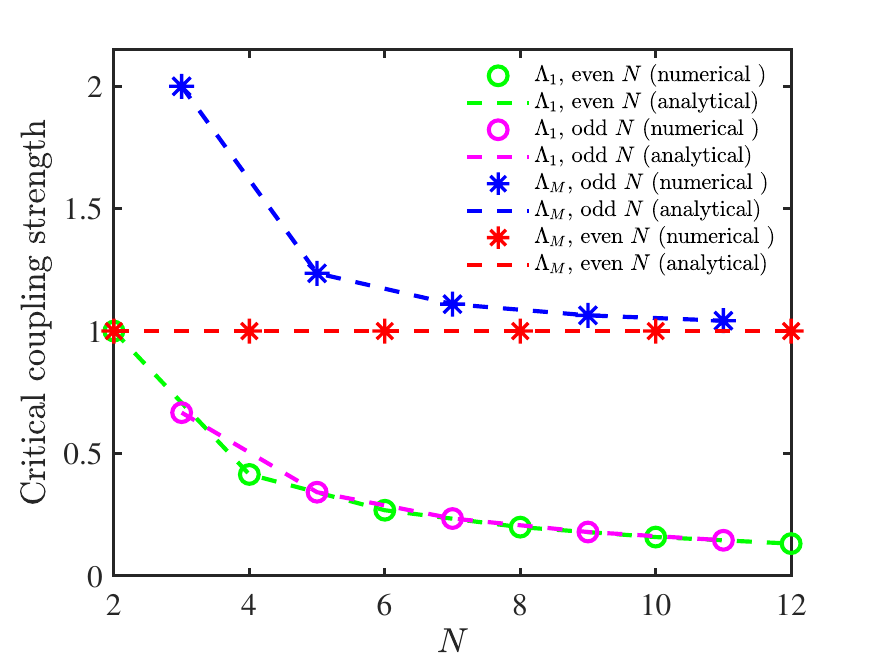}
    \caption{Critical interaction strengths $\Lambda_1$ and $\Lambda_M$ as functions of the chain length $N$. Here, $\Lambda_1$ marks the first level crossing, whereas $\Lambda_M$ marks the crossing at which the fully polarized ground state is replaced by the lowest one-excitation state. Green open circles and the green dashed curve show, respectively, the numerical and analytical values of $\Lambda_1$ for even $N$, while magenta open circles and the magenta dashed curve show the corresponding results for odd $N$. Blue pentagrams and the blue dashed curve show, respectively, the numerical and analytical values of $\Lambda_M$ for odd $N$, whereas red asterisks and the red dashed curve show the corresponding results for even $N$. We set $\omega_0=1$.}
    \label{5555}
\end{figure}

\begin{figure}[!t]
    \centering
    \includegraphics[width=3.3in]{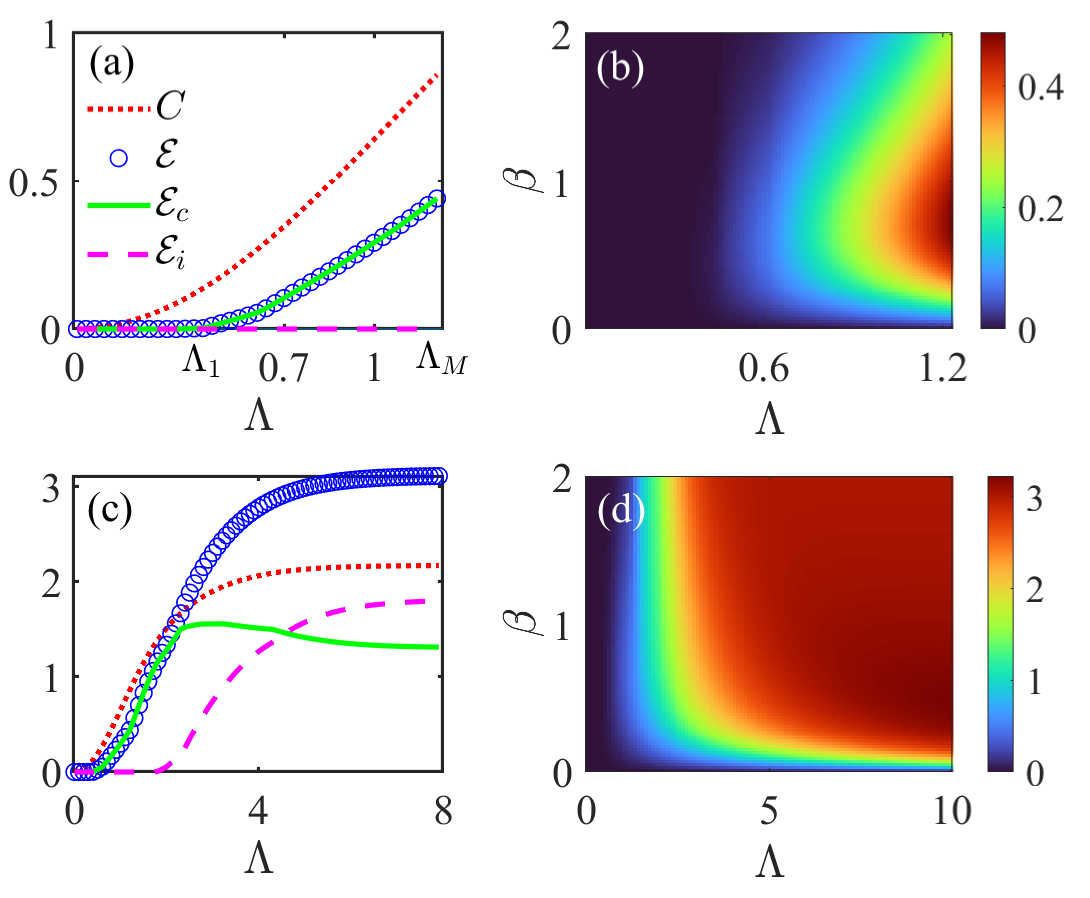}
     \caption{Ergotropy and coherence in the $N$-TLS XX spin chain with $N=5$. The total ergotropy, its coherent and incoherent contributions, and the quantum coherence $C$ defined in Eq.~\eqref{C} are shown as functions of the coupling strength $\Lambda$ at fixed inverse temperature $\beta=1$ in panels (a) and (c). Panels (b) and (d) show the total ergotropy as a function of both $\Lambda$ and $\beta$. The range of coupling strength in (a) and (b) is $0\leq \Lambda \leq \Lambda_M$, while the range in (c) and (d) encompasses both the weak- and strong-coupling regimes. The onsite energy is set to $\omega_0=1$.
 }
  \label{4444}
\end{figure}

For $\beta>0$, the Gibbs weights at $\Lambda=0$ are nonincreasing with increasing bare energy, so the thermal state is passive with respect to $H_0$. For $0<\Lambda<\Lambda_1$, no two lines belonging to different excitation subspaces have crossed, and this ordering remains unchanged. Rotations inside a fixed $n$ subspace do not alter the conclusion because all vectors in that subspace have the same bare energy $(-N+2n)\omega_0$. At $\Lambda=\Lambda_1$, the crossing lines have equal Gibbs weights, whereas immediately above it the higher-$H_0$ subspace acquires the larger population. This population inversion makes the Gibbs state of $H_S$ non-passive, with strictly positive ergotropy for any finite $\beta>0$. Thus, $\Lambda_1$ is the activation boundary. The ergotropy is positive for $\Lambda>\Lambda_1$, while at $\Lambda=\Lambda_1$ the crossing lines contribute no ergotropy because their populations are equal.

In the large-$N$ limit, $\Lambda_1\sim\pi\omega_0/(2N)$, whereas $\Lambda_M\longrightarrow\omega_0$.
The $1/N$ behavior of $\Lambda_1$ is a many-body spectral effect. Near half filling, the maximal difference in interaction slopes between neighboring excitation states grows linearly with $N$, while their bare-energy separation remains $2\omega_0$. By contrast, $\Lambda_M$ is set by the lowest one-particle mode and remains finite. Increasing $N$ therefore makes the first ergotropy-activating spectral rearrangement accessible at progressively weaker coupling, but the coupling required to reorder the ground state remains finite. See Fig.~\ref{5555} for numerical illustrations. 

For $N>2$, in the interval $\Lambda_1<\Lambda<\Lambda_M$, the level inversion occurs only among thermally populated excited states. It yields positive ergotropy at finite temperature, but its contribution is exponentially suppressed as $\beta\to\infty$, because the Gibbs state then approaches the ground state. Once $\Lambda>\Lambda_M$, the ground state has been replaced and non-passivity can survive the zero-temperature limit. For the $N=5$ example in Fig.~\ref{4444}, the two analytical thresholds are $\Lambda_1/\omega_0=\tan(\pi/10)\sec(\pi/10)\simeq0.34164$ and $\Lambda_M/\omega_0=\sec(\pi/5)\simeq1.23607$. Because $H_0$ is degenerate within each fixed-excitation subspace, we fix the local product basis when applying the dephasing map $\Delta$. The coherent and incoherent contributions reported below are defined with respect to this convention. The numerical decomposition in Fig.~\ref{4444} shows that, for the parameters plotted, the strong-coupling regime supports both contributions, which approach well-defined asymptotic values as $\Lambda$ increases.
\endgroup
\hypersetup{linkcolor=blue,citecolor=blue}

\section{Conclusion}

We have investigated ergotropy generation in dissipatively charged quantum batteries and analyzed its coherent and incoherent components in resonant, detuned, and many-body settings. In the two-qubit case, the resonant setup exhibits a clear activation threshold: extractable work appears only beyond the critical coupling $\Lambda_M$, with the coherent contribution saturating at $\omega/3$ in the low-temperature regime and approaching the strong-coupling equipartition limit $\mathcal{E}_c=\mathcal{E}_i=\omega/2$. By contrast, detuning lowers the activation threshold and enables ergotropy already in the weak-coupling regime, including parameter regions where work extraction is purely coherent. We further show that quantum coherence, although necessary for coherent work extraction, is not by itself sufficient to guarantee nonzero coherent ergotropy; non-passivity must also be induced, either by frequency detuning or by driving the interaction beyond the relevant level-crossing threshold.

Extending the analysis to $N$-TLS XX spin chains, we show that increasing system size lowers the coupling strength of the first relevant level crossing and thereby facilitates ergotropy activation. The many-body results therefore support the same physical picture identified in the two-qubit case: interaction-induced spectral rearrangement, together with coherence generation, governs the onset of extractable work. Taken together, these results identify frequency detuning, coupling strength, and system size as useful control parameters for optimizing work extraction in dissipatively charged quantum batteries.

These results complement previous studies of dissipative charging by showing how interaction-driven thermalization can activate not only stored energy, but also its extractable coherent and incoherent work components~\cite{barra2019dissipative,mayo2022collective,pokhrel2025large}. In particular, our results sharpen the relation between coherence and work extraction discussed in Refs.~\cite{PhysRevLett.125.180603,Shi2022Entanglement} by showing that coherence alone does not guarantee nonzero coherent ergotropy in dissipatively charged batteries. Given the relevance of exchange-coupled spin models to platforms such as superconducting qubits and Rydberg-atom arrays, our results may also provide useful guidance for experimentally exploring ergotropy activation under dissipative charging conditions~\cite{Nguyen2024,Zhou2021,Lippe2021}.
\begingroup

More concretely, the parameters $\omega_j$ in our Hamiltonian should be interpreted experimentally as tunable local $z$ fields or detunings in a rotating or Floquet frame, rather than as the bare qubit transition frequencies, while $\Lambda$ corresponds to an engineered XY exchange coupling. These two ingredients have already been realized as tunable onsite potentials and $(X_iX_j+Y_iY_j)/2$ exchange couplings in superconducting quantum processors, while resonant dipole--dipole interactions provide the corresponding XY exchange in Rydberg-atom arrays~\cite{Andersen2025,Barredo2015}. Following a quench of the exchange coupling, two-qubit state tomography could reconstruct the charged state and hence its ergotropy components. The main additional challenge for an experimental implementation is controlled thermalization to the global Gibbs state of the interacting Hamiltonian.
\endgroup

\appendix
\begingroup

\section{Passive-state derivation of the two-TLS ergotropy}
\label{app:passive_derivation}

We derive Eqs.~\eqref{e}, \eqref{ei}, and \eqref{ec} directly from the passive-state ordering rule. The four eigenenergies of $H_0$ are $\{\omega,\delta,-\delta,-\omega\}$ and hence, in nondecreasing order, $-\omega,-|\delta|,|\delta|,\omega$. Because the eigenvalues of $\rho=e^{-\beta H_S}/Z$ decrease with the eigenenergies of $H_S$, their nonincreasing ordering can be represented by the following unnormalized Gibbs weights:
\begin{equation}
\begin{cases}
(e^{\beta\omega},e^{\beta\kappa},e^{-\beta\kappa},e^{-\beta\omega}), & \kappa\leq\omega,\\
(e^{\beta\kappa},e^{\beta\omega},e^{-\beta\omega},e^{-\beta\kappa}), & \kappa>\omega.
\end{cases}
\label{eq:app_rho_order}
\end{equation}
The two orderings meet at $\kappa=\omega$, which is equivalent to $\Lambda=\Lambda_M$ because $\kappa^2-\omega^2=4(\Lambda^2-\omega_1\omega_2)$.
Assigning the corresponding normalized populations to the ordered bare energies gives
\begin{equation}
E(\rho_p)=-\frac{2}{Z}
\begin{cases}
\omega\sinh(\beta\omega)+|\delta|\sinh(\beta\kappa), & \Lambda\leq\Lambda_M,\\
\omega\sinh(\beta\kappa)+|\delta|\sinh(\beta\omega), & \Lambda>\Lambda_M.
\end{cases}
\label{eq:app_passive_total}
\end{equation}
Subtracting this result from $E(\rho)=-2[\delta^2\sinh(\beta\kappa)+\kappa\omega\sinh(\beta\omega)]/(\kappa Z)$ yields Eq.~\eqref{e}.

For the dephased state in Eq.~\eqref{rhodiag}, the populations associated with the bare energies $-|\delta|$ and $|\delta|$ are, respectively,
\begin{align}
\pi_-&=\max\{\pi^{22},\pi^{33}\}
=\frac{\cosh(\beta\kappa)+(|\delta|/\kappa)\sinh(\beta\kappa)}{Z},\nonumber\\
\pi_+&=\min\{\pi^{22},\pi^{33}\}
=\frac{\cosh(\beta\kappa)-(|\delta|/\kappa)\sinh(\beta\kappa)}{Z}.
\label{eq:app_dephased_populations}
\end{align}
The only ordering changes relevant to passivity are therefore the crossings $\pi^{22}=\pi^{44}$ and $\pi^{33}=\pi^{44}$ in Eqs.~\eqref{A} and \eqref{B}. Irrespective of the sign of $\delta$, the larger population $\pi_-$ crosses $\pi^{44}$ at $\Lambda_{\min}$, whereas $\pi_+$ crosses it at $\Lambda_{\max}$. Thus, listing the populations assigned to the ordered energies $(-\omega,-|\delta|,|\delta|,\omega)$, the passive state has the ordering
\begin{equation}
\begin{cases}
(\pi^{44},\pi_-,\pi_+,\pi^{11}), & \Lambda\leq\Lambda_{\min},\\
(\pi_-,\pi^{44},\pi_+,\pi^{11}), & \Lambda_{\min}<\Lambda\leq\Lambda_{\max},\\
(\pi_-,\pi_+,\pi^{44},\pi^{11}), & \Lambda>\Lambda_{\max}.
\end{cases}
\label{eq:app_pi_passive}
\end{equation}
Consequently,
\begin{equation}
\begin{aligned}
\mathcal E_i&=E(\pi)-E(\pi_p)\\
&=
\begin{cases}
0, & \Lambda\leq\Lambda_{\min},\\
(\omega-|\delta|)(\pi_--\pi^{44}), & \Lambda_{\min}<\Lambda\leq\Lambda_{\max},\\
\begin{gathered}
(\omega-|\delta|)\pi_-+2|\delta|\pi_{+}\\
{}-(\omega+|\delta|)\pi^{44}
\end{gathered}, & \Lambda>\Lambda_{\max}.
\end{cases}
\end{aligned}
\label{eq:app_Ei_compact}
\end{equation}
Substitution of Eq.~\eqref{eq:app_dephased_populations} into Eq.~\eqref{eq:app_Ei_compact} gives Eq.~\eqref{ei}. Finally, $\mathcal E_c=\mathcal E-\mathcal E_i$, together with $\Lambda_M<\Lambda_{\min}$, generates the four intervals in Eq.~\eqref{ec}. Direct subtraction then gives the four corresponding expressions without requiring any additional population ordering.

\section{Analytical derivation of \texorpdfstring{$\Lambda_M$ and $\Lambda_1$}{LambdaM and Lambda1} for the periodic XX chain}
\label{app:xx_thresholds}

This Appendix provides a derivation of the two critical couplings $\Lambda_1$ and $\Lambda_M$ in Sec.~\ref{N TLS}. The physical reason why the first crossing activates ergotropy is discussed in the main text. Here we establish the many-body spectrum, determine its extremal slopes, and use them to prove the formulas for $\Lambda_M$ and $\Lambda_1$.

\subsection{Excitation spectrum and extremal interaction slopes}
\label{app:excitation_slopes}

Introduce the conserved excitation-number operator
\begin{equation}
\hat n=\sum_{j=1}^{N}\frac{\sigma_j^z+I}{2}.
\end{equation}
Since $[H_S,\hat n]=0$, the Hilbert space decomposes as $\mathcal H=\bigoplus_{n=0}^{N}\mathcal H_n$. In the subspace with $n$ excitations,
\begin{equation}
H_S\big|_{\mathcal H_n}=(-N+2n)\omega_0 I+\Lambda V_n .
\label{eq:app_subspace}
\end{equation}
If $V_n\ket{\phi_{n,\alpha}}=\mu_{n,\alpha}\ket{\phi_{n,\alpha}}$, the corresponding eigenvalue is
\begin{equation}
E_{n,\alpha}(\Lambda)
=(-N+2n)\omega_0+\mu_{n,\alpha}\Lambda ,
\label{eq:app_affine}
\end{equation}
which proves Eq.~\eqref{eq:N_affine}. Two eigenvalues in $n<m$ cross at positive coupling only if $\mu_{n,\alpha}>\mu_{m,\beta}$, in which case
\begin{equation}
\Lambda_{(n,\alpha),(m,\beta)}
=\frac{2(m-n)\omega_0}
{\mu_{n,\alpha}-\mu_{m,\beta}} .
\label{eq:app_general_crossing}
\end{equation}

The Jordan--Wigner transformation followed by a Fourier transformation gives the free-fermion representation \cite{franchini2008renyi}
\begin{equation}
H_S=-N\omega_0+
\sum_{k\in K_n}
\left(2\omega_0+2\Lambda\cos k\right)c_k^\dagger c_k ,
\label{eq:app_JW}
\end{equation}
where the allowed momentum grid depends on the fermion-number parity,
\begin{equation}
K_n=
\begin{cases}
\left\{\dfrac{2\pi m}{N}\,\middle|\,m=0,\ldots,N-1\right\}, & n\ \text{odd},\\[2mm]
\left\{\dfrac{2\pi(m+\tfrac12)}{N}\,\middle|\,m=0,\ldots,N-1\right\}, & n\ \text{even}.
\end{cases}
\label{eq:app_grid}
\end{equation}
Momenta are understood modulo $2\pi$. For a configuration $Q\subset K_n$ of $n$ occupied modes, the interaction slope is
\begin{equation}
\mu(Q)=2\sum_{k\in Q}\cos k .
\label{eq:app_slope}
\end{equation}

Set $\theta=\pi/N$. The maximum (minimum) slope is obtained by occupying the allowed momenta closest to $0$ ($\pi$). Using
\begin{equation}
\sum_{r=0}^{q-1}\cos(\alpha+2r\theta)
=\frac{\sin(q\theta)}{\sin\theta}
\cos\!\left[\alpha+(q-1)\theta\right],
\label{eq:app_sum}
\end{equation}
and matching the periodic (antiperiodic) grid to odd (even) $n$ gives
\begin{equation}
\mu_n^{\max}=A_n,
\qquad
A_n=2\frac{\sin(n\theta)}{\sin\theta}.
\label{eq:app_An}
\end{equation}

Because the cosine sum over either complete grid vanishes, the minimum $n$-particle sum is minus the maximum sum of its $(N-n)$-mode complement. The complement has the same parity as $n$ for even $N$ and the opposite parity for odd $N$, yielding
\begin{equation}
\mu_n^{\min}=
\begin{cases}
-A_n, & N\ \text{even},\\
-\cos\theta\,A_n, & N\ \text{odd}.
\end{cases}
\label{eq:app_min_slopes}
\end{equation}
These extremal slopes are sufficient to determine both thresholds.

\subsection{Ground state crossing threshold \texorpdfstring{$\Lambda_M$}{LambdaM}}
\label{app:LM}

The fully polarized level has energy $E_0=-N\omega_0$ and zero slope, whereas the lowest level in the subspace $\mathcal H_n$ has energy
\begin{equation}
E_n^{\min}=-N\omega_0+2n\omega_0
+\Lambda\mu_n^{\min}.
\end{equation}
Its crossing with $E_0$ occurs at
\begin{equation}
\Lambda_{M,n}=
\begin{cases}
\dfrac{n\omega_0\sin\theta}{\sin(n\theta)}, & N\ \text{even},\\[3mm]
\dfrac{n\omega_0\sin\theta}{\cos\theta\,\sin(n\theta)}, & N\ \text{odd}.
\end{cases}
\label{eq:app_LMn}
\end{equation}
For $1\leq n\leq N-1$, $\sin(n\theta)\leq n\sin\theta$, with equality at $n=1$. Thus the earliest crossing of the fully polarized level is
\begin{equation}
\Lambda_M=
\begin{cases}
\omega_0, & N\ \text{even},\\[1mm]
\omega_0\sec\theta, & N\ \text{odd},
\end{cases}
\qquad \theta=\frac{\pi}{N},
\label{eq:app_LM_final}
\end{equation}
which proves Eq.~\eqref{eq:N_LM}. The parity factor has a simple origin: the one-particle grid contains $k=\pi$ for even $N$, whereas the closest modes for odd $N$ are $k=\pi\pm\pi/N$.

\subsection{Earliest positive crossing \texorpdfstring{$\Lambda_1$}{Lambda1}}
\label{app:L1}

Define the largest slope-closing rate between neighboring excitation subspaces as
\begin{equation}
D_N=\max_{n,\alpha,\beta}
\left(\mu_{n,\alpha}-\mu_{n+1,\beta}\right)>0 .
\label{eq:app_D_def}
\end{equation}
For any two excitation subspaces separated by $r=m-n\geq1$, inserting arbitrary intermediate levels and telescoping gives $\mu_{n,\alpha}-\mu_{m,\beta}\leq rD_N$. Equation~\eqref{eq:app_general_crossing} therefore implies
\begin{equation}
\Lambda_{(n,\alpha),(m,\beta)}
\geq\frac{2r\omega_0}{rD_N}
=\frac{2\omega_0}{D_N}.
\end{equation}
An adjacent pair attaining the maximum in Eq.~\eqref{eq:app_D_def} saturates this bound, hence
\begin{equation}
\Lambda_1=\frac{2\omega_0}{D_N}.
\label{eq:app_L1_D}
\end{equation}

For even $N=2L$, the largest slope difference between $\mathcal H_n$ and $\mathcal H_{n+1}$ is
\begin{equation}
D_n=A_n+A_{n+1}
=\frac{2\sin[(n+\tfrac12)\theta]}{\sin(\theta/2)}.
\end{equation}
It is maximal at the pair(s) nearest half filling, giving $D_N^{\rm even}=2\cot(\theta/2)$. For odd $N=2L+1$,
\begin{equation}
D_n=A_n+\cos\theta A_{n+1}.
\end{equation}
The sequence $A_n$ is symmetric and unimodal, with $A_L=A_{L+1}=1/\sin(\theta/2)$, hence the central pair gives $D_N^{\rm odd}=2\cot(\theta/2)\cos(\theta/2)$. Substitution into Eq.~\eqref{eq:app_L1_D} yields
\begin{equation}
\Lambda_1=
\begin{cases}
\omega_0\tan\!\left(\dfrac{\pi}{2N}\right), & N\ \text{even},\\[2mm]
\omega_0\tan\!\left(\dfrac{\pi}{2N}\right)
\sec\!\left(\dfrac{\pi}{2N}\right), & N\ \text{odd},
\end{cases}
\label{eq:app_L1_final}
\end{equation}
which proves Eq.~\eqref{eq:N_L1}.

The numerical markers in Fig.~\ref{5555} were obtained independently by diagonalizing each fixed-excitation matrix $V_n$ once and evaluating Eq.~\eqref{eq:app_general_crossing}. Therefore, they contain no $\Lambda$-grid error. For $N=2$, the periodic sum counts the unique bond twice, and the convention used here gives $\Lambda_1=\Lambda_M=\omega_0$, whereas a single-bond two-qubit convention requires the corresponding rescaling of $\Lambda$.
\endgroup
\hypersetup{linkcolor=blue,citecolor=blue}

\begin{acknowledgments}
This work was supported by the NSFC (Grants No.12275215, No.12305028, and No.12574300), and the Youth Innovation Team of Shaanxi Universities. KZ is supported by the China Postdoctoral Science Foundation under Grant Number 2025M773421, Shaanxi Province Postdoctoral Science Foundation under Grant Number 2025BSHYDZZ017, and Scientific Research Program Funded by Education Department of Shaanxi Provincial Government (Program No.24JP186 and No.25JP183).
\end{acknowledgments}

%

\end{document}